\documentclass[%
reprint,
 amsmath,amssymb,
 aps,
floatfix,
]{revtex4-2}
\UseRawInputEncoding
\usepackage{graphicx}
\usepackage{dcolumn}
\usepackage{bm}
\usepackage{xcolor}
\usepackage[caption = false]{subfig}
\usepackage[utf8]{}
\usepackage{float}
\usepackage{amsmath}
\usepackage{amsfonts}
\usepackage{graphicx}

\begin{document}

\title{Critical Dynamics of the Antiferromagnetic $O(3)$ Nonlinear Sigma Model with Conserved Magnetization}

\author{Louie Hong Yao}
\author{Uwe C. T\"auber}
\email{tauber@vt.edu}

\affiliation{Department of Physics \& Center for Soft Matter and Biological Physics, MC 0435,
		Robeson Hall, 850 West Campus Drive, Virginia Tech, Blacksburg, VA 24061, USA}

\date{\today}

\begin{abstract}
We study the near-equilibrium critical dynamics of the $O(3)$ nonlinear sigma model describing isotropic antiferromagnets
with non-conserved order parameter reversibly coupled to the conserved total magnetization. 
To calculate response and correlation functions, we set up a description in terms of Langevin stochastic equations of motion,
and their corresponding Janssen--De~Dominicis response functional. 
We find that in equilibrium, the dynamics is well-separated from the statics, at least to one-loop order in a perturbative
treatment with respect to the static and dynamical nonlinearities. 
Since the static nonlinear sigma model must be analyzed in a dimensional $d = 2 + \varepsilon$ expansion about its lower 
critical dimension $d_\textrm{lc} = 2$, whereas the dynamical mode-coupling terms are governed by the upper critical 
dimension $d_c = 4$, a simultaneous perturbative dimensional expansion is not feasible, and the reversible critical dynamics 
for this model cannot be accessed at the static critical renormalization group fixed point. 
However, in the coexistence limit addressing the long-wavelength properties of the low-temperature ordered phase, we can
perform an $\epsilon = 4 - d$ expansion near $d_c$.
This yields anomalous scaling features induced by the massless Goldstone modes, namely sub-diffusive relaxation for the 
conserved magnetization density with asymptotic scaling exponent $z_\Gamma = d - 2$ which may be observable in neutron 
scattering experiments.
Intriguingly, if initialized near the critical point, the renormalization group flow for the effective dynamical exponents recovers
their universal critical values $z_c = d / 2$ in an intermediate crossover region.
\end{abstract}



\maketitle

\section{Introduction}

The universal scaling behavior near critical points and in low-temperature ordered phases with spontaneously broken
continuous symmetries has been studied extensively in both equilibrium thermal as well as quantum systems since the 1960s.
Many different powerful methods have been developed, most prominently Monte Carlo computer simulations and 
renormalization group (RG) analysis, carried out either perturbatively analytically (see, for example,
Refs.~\cite{wilson1972critical,wilson1974renormalization,wilson1983renormalization}) or non-perturbatively numerically 
(e.g., Refs.~\cite{hasselmann2004critical,stokic2010functional,litim2011ising}); we also note the recently developed 
``conformal bootstrap'' approach \cite{el2012solving,poland2016conformal,poland2019conformal}. 

Yet for systems driven away from thermal equilibrium, more complicated and intriguing issues arise. 
It is well-known that systems in the same static equilibrium universality class may in fact display distinct universal dynamic 
scaling features. 
$O(n)$-symmetric magnetic systems constitute a paradigmatic example: 
The associated order parameter may either freely relax to equilibrium or be constrained by a strict conservation law for the
total magnetization, whence the magnetization density relaxes diffusively.
These two different variants are described by the relaxational models A and B in Hohenberg and Halperin's classification
\cite{hohenberg1977theory}. 
In addition to purely relaxational kinetics, the order parameter may be reversibly dynamically coupled to other conserved
modes (c.f., models E, G, and J).

To distinguish different dynamic universality classes near equilibrium, a distinct critical exponent, the dynamic scaling 
exponent $z \geq 1$, was introduced to capture the phenomenon of critical slowing down of a critical system's kinetics
\cite{ferrell1967dispersion,halperin1969scaling}.
As this term implies, near a critical point the order parameter is governed by slow algebraic relaxation rather than the
standard exponential decay with finite characteristic relaxation time $t_r$. 
This power law behavior can be described by a scaling law for the relaxation time $t_r \sim \xi^z$, where 
$\xi \sim |T - T_c|^{-\nu}$ ($\nu > 0$) denotes the diverging correlation length of the system; or equivalently by the 
leading wavevector dependence of the order parameter damping coefficient $D(q) \sim |q|^z$ that determines the peak linewidth in the associated scattering cross section.

One important historical approach to theoretically classify dynamic universal behavior was mode-coupling theory, wherein
reversible dynamical coupling terms were added to the relaxation dynamics, and the resulting stochastic equations of motion 
subsequently solved by means of a Hartree-Fock like self-consistent factorization approximation. 
This methodology was originally proposed by Fixman for binary fluids \cite{fixman1962viscosity}, later reformulated by
Kadanoff and Swift \cite{kadanoff1968transport} and Kawasaki, who also extended this treatment to magnetic systems
\cite{kawasaki1967anomalous,kawasaki1970kinetic,gunton1976renormalization}. 
In a theory incorporating dynamical mode-coupling terms, the order parameter is coupled to other slow hydrodynamic 
modes, which originate from conservation laws, during the system's relaxation. 
For example, in the still purely relaxational models C / D, a static coupling term between a real scalar field and the order
parameter is introduced into the Hamiltonian to keep the total energy conserved 
\cite{hohenberg1977theory,vasil2004field,folk2006critical,tauber2014critical}.  
Furthermore, when the conserved modes correspond to specific order parameter symmetries, additional reversible dynamical 
terms can be introduced via the Lie algebra Poisson brackets between the order parameter and the conserved modes \cite{halperin1974renormalization,ma1974critical,sasvari1975hydrodynamics,halperin1980erratum,sasvari1977dynamic, de1978field,folk2006critical,tauber2014critical}. 

In this work, we utilize the field-theoretic perturbative dynamical renormalization group approach to study the 
near-equilibrium relaxation dynamics of the antiferromagnetic $O(3)$ nonlinear sigma model with conserved total
magnetization.
We specifically account for the reversible hydrodynamic mode couplings between the conserved magnetization density fields
and the non-conserved staggered magnetization serving as the order parameter for the continuous para- to antiferromagnetic
phase transition.
For magnetic systems, these purely dynamical contributions originate from the microscopic spin precession in the local 
effective field.
In an antiferromagnetically ordered phase, in contrast with purely relaxational kinetics, the mode coupling of the staggered 
magnetization to the conserved standard magnetization density causes the emergence of propagating spin wave modes 
(magnons) with linear dispersion $\omega(q) = c \, |q|$ and quadratic wave vector dependence of the damping 
$D(q) \sim q^2$ [see Eq.~(\ref{dispersion1}) below].
At $T = T_c$, strong order parameter fluctuations induce the critical damping $D(q) \sim \lambda(q) \sim |q|^{z_c}$ for
both the order parameter and the magnetization densities, with the well-stablished dynamic critical exponent $z_c = d / 2$ in 
dimensions $2 < d < d_c = 4$, which represents the upper critical dimension for this model G dynamical universality class
\cite{hohenberg1977theory,folk2006critical,tauber2014critical}.
This exponent value differs markedly from the corresponding mean-field values $z_0 = 2$ and the purely relaxational model
A result $z_c = 2 + c' \, \eta$, where $\eta$ denotes the static Fisher critical exponent $\eta$ and $c'$ is a universal constant.
These classical results have been amply confirmed experimentally, e.g., perhaps most convincingly in neutron scattering 
experiments for the isotropic antiferromagnet RbMnF$_3$ \cite{coldea1998critical} as well as in detailed numerical 
simulations, for example in Refs.~\cite{bunker1996critical,tsai2003critical,nandi2020critical}.

In thermal equilibrium, the $O(n)$ nonlinear sigma model represents a paradigmatic field theory to describe ferromagnetic or
antiferromagnetic systems exhibiting rotational symmetry in order parameter space 
\cite{brezin1976renormalization,chakravarty1989two}. 
It exhibits spontaneous symmetry breaking  and a continuous phase transition at a critical temperature $T_c$, and its static
thermodynamic critical properties are well-established to be in the same universality class as the $O(n)$-symmetric 
Landau--Ginzburg--Wilson $\phi^4$ theory in equilibrium.
Out of equilibrium, the $O(n)$ nonlinear sigma model has to our knowledge only been studied under the restrictive 
assumption of purely relaxational (model A) dynamics. 
In 1980, Bausch, Janssen, and Yamazaki calculated the associated dynamical critical exponent $z$ and compared it with the
corresponding scaling exponent of the relaxational $O(n)$-symmetric Landau--Ginzburg--Wilson theory 
\cite{bausch1980nonlinear}.
Only comparatively recently in 2006, Fedorenko and Trimper evaluated the critical aging scaling behavior for the purely 
relaxational $O(n)$ nonlinear sigma model with non-conserved order parameter \cite{fedorenko2006critical}; not 
surprisingly, it belongs to the model A universality class. 
Although the $O(n)$-symmetric Landau--Ginzburg--Wilson model with mode-coupling terms have been widely studied in the 
literature (for overviews, see Refs.~\cite{folk2006critical,tauber2014critical}), the $O(n)$ nonlinear sigma model with 
non-dissipative reversible mode-mode couplings appears to have not drawn similar attention.

In this paper, we couple the three-component non-conserved staggered magnetization order parameter of the nonlinear 
sigma model to the conserved magnetization density vector field. 
We employ coupled Langevin-type stochastic partial differential equations to describe the dynamics of this system and 
construct the associated Janssen--De~Dominicis functional \cite{janssen1976lagrangean,de1975field,brezin1975field}. 
In the resulting field theory framework, we apply the standard transverse fluctuation loop expansion to perturbatively 
compute the model's dynamical response functions to the first nontrivial order in the nonlinear couplings. 
Indeed, to one-loop order, the dynamics cleanly separates from the statics, and in the low-frequency limit $\omega \to 0$, 
we properly recover the static susceptibilities. 
However, it turns out that the ordinary dimensional $d = 2 + \varepsilon$ expansion about the lower critical dimension 
$d_\textrm{lc} = 2$ that captures the nonlinear sigma model's universal static critical properties is not capable to access the 
critical dynamics in the presence of reversible mode couplings.

Hence we turn our attention to the ordered-phase or coexistence fixed point; to this end, we tune the effective temperature to
zero and employ an $\epsilon = 4 - d$ expansion scheme near the dynamical upper critical dimension $d_c = 4$ to retrieve 
the universal dynamical scaling behavior in the coexistence regime. 
Solving the one-loop RG flow equations, we find that the dynamic scaling exponents for both the order parameter and for the 
magnetization density flow to the same fixed point values as in the Sasv\'ari--Schwabl--Sz\'epfalusy (SSS) model 
\cite{tauberdissertation}. 
Hence we obtain Gaussian dynamical scaling exponents $z_D = 2$ and $z_\lambda = 2$ for the order parameter and 
transverse magnetization components, implying a mean-field form for the ensuing spin wave dispersion, but our results 
remarkably yield anomalous sub-diffusive relaxation for the longitudinal magnetization density, described by 
$z_\Gamma = d - 2$.
This result indicates that the $O(n)$ nonlinear sigma model with conserved total magnetization belongs to the same 
dynamical universality class as the SSS model. 
Intriguingly, although the critical regime is conceptionally not approachable with an $\epsilon$ expansion near the upper
critical dimension $d_c = 4$, the numerical solutions of the near-critical RG flow equations initially approach the known 
critical values, recovering the dynamic critical exponent $z_c = d / 2$, before they ultimately cross over to their asymptotic 
scaling behavior at the ordered-phase coexistence fixed point.

The paper is organized as follows: 
In the following section, we introduce the near-equilibrium dynamics of the antiferromagnetic nonlinear sigma model with
conserved total magnetization, properly incorporating the crucial non-dissipative mode coupling terms. 
We formulate the coupled Langevin stochastic equations for the order parameter and the magnetization density, construct the 
corresponding Janssen--De~Dominicis functional, and utilize this formalism to derive the explicit one-loop expressions for the
dynamic response functions. 
In Section III, we discuss the ensuing scaling behavior of the model near its RG fixed points. 
We argue that the critical fixed point is not reachable in this perturbative approach. 
Subsequently we present a thorough scaling analysis via the dimensionally regularized perturbative RG method near the 
ordered phase or coexistence fixed point. 
We numerically solve the resulting RG flow equations and compare our findings with the SSS model, and discuss relevant 
experimental consequences.
We conclude with a brief summary and outlook.
The Appendix contains additional pertinent technical details for the one-loop perturbative analysis.
It also addresses the feasibility of the critical dynamics of the nonlinear sigma model with conserved total energy as well as 
a conserved order parameter field.

\section{Nonlinear sigma model with conserved total magnetization}

In this section, we introduce the near-equilibrium dynamics of the $O(3)$-symmetric antiferromagnetic nonlinear sigma 
model with conserved total magnetization.
Its time evolution subject to random thermal fluctuations is described by appropriate coupled generalized Langevin equations, 
and the corresponding Janssen--De~Dominicis response field theory functional that encodes this stochastic nonlinear 
dynamics. 
We employ dynamic perturbation theory to derive explicit results for the dynamic response functions to one-loop order in the
fluctuation expansion with respect to the transverse modes.
The ensuing dynamic scaling behavior will be discussed in the next section.

\subsection{Dynamics with conserved total magnetization}

We consider a rotationally symmetric three-component antiferromagnet described by the $O(3)$ nonlinear sigma model with 
the coarse-grained mesoscopic Hamiltonian \cite{brezin1976renormalization}
\begin{equation}
    \mathcal{H}[{\vec n}] = \int\mathrm{d}^dx \left[ \frac{\rho}{2} \Big( \nabla\vec{n}(x) \Big)^2 - h \, \sigma \right] ,
    \label{statichamiltonian}
\end{equation}
where $\vec{n}(x) = \big( \vec{\pi}(x),\sigma(x) \big)$ denotes the staggered magnetization vector that serves as order 
parameter for an antiferromagnet, and the constraint $\vec{n}^2 = 1$ is imposed at all locations $x$. 
$\rho$ represents the order parameter stiffness and $h$ is a (fictitious) external conjugate thermodynamic field pointed along 
the direction of $\sigma$. 
(Note that positions $x$ and later, the corresponding wavevectors and momenta $p$, are $d$-dimensional spatial vectors. 
We drop their vector labels in order to not confuse them with the three-component vector fields in order parameter space.) 

To describe the interaction of the antiferromagnet with the conserved total magnetization, we introduce the magnetization 
density $\vec{M}(x,t)$ in the Hamiltonian (\ref{statichamiltonian}),
\begin{equation}
    \mathcal{H}[{\vec M},{\vec n}] = \int\mathrm{d}^dx \left[ \frac{1}{2\chi} \vec{M}(x)^2 
    + \frac{\rho}{2} \Big( \nabla\vec{n}(x) \Big)^2 - h \, \sigma \right] ,
    \label{Hamiltonian} 
\end{equation}
where $\chi$ denotes the static magnetic susceptibility. 
The conservation of the total magnetization is intimately related to the underlying $O(3)$ invariance for the order parameter.
Indeed, ${\vec M}$ is the generator of the rotational symmetry, and hence satisfies the standard angular momentum Lie
algebra Poisson brackets
\begin{equation}
\begin{aligned}
    \{M_\alpha(x), M_\beta(y)\} &= - \epsilon_{\alpha\beta\gamma} \, M_\gamma(x) \, \delta^{(d)}(x-y) \ , \\
    \{M_\alpha(x), n_\beta(y)\} &= - \epsilon_{\alpha\beta\gamma} \, n_\gamma(x) \, \delta^{(d)}(x-y) \ ,
\end{aligned}
\label{lie bracket}
\end{equation}
where $\alpha, \beta, \gamma =1, 2, 3$, and Einstein's summation rule over repeated indices is implemented (unless 
otherwise specified). 

Generalized Langevin equations for ``slow" modes $\phi$ with respect to an effective Hamiltonian $\mathcal{H}[\phi]$ take 
the form
\begin{equation}
    \dot{\phi} = \{\phi,\mathcal{H}\} - D(i\nabla)^{a} \frac{\delta \mathcal{H}}{\delta \phi} + \eta \ ,
\end{equation}
where $a=0,2$ corresponds to nonconserved and conserved relaxation dynamics, respectively, $D$ denotes the relaxation rate or diffusion constant, and $\eta$ represents Gaussian stochastic noise. 
With the Hamiltonian (\ref{Hamiltonian}) and Poisson brackets (\ref{lie bracket}), the following coupled Langevin dynamics
of the antiferromagnetic order parameter $\vec{n}$ and the rescaled conserved magnetization density 
$\vec{m} = \vec{M} / \sqrt{\rho\chi}$ ensue \cite{hohenberg1977theory,tauber2014critical}, 
\begin{equation}
\begin{aligned}
    \dot{\vec{m}} &=  c \, \vec{n} \times \nabla^ 2 \vec{n} + \lambda \, \nabla^2 \vec{m} + \vec{\eta} \ , \\
    \dot{\vec{n}}  &= -c \, \vec{n} \times \vec{m} + D \left( \nabla^2 \vec{n} + \tilde{h} \right) + \vec{\zeta} \ .
\end{aligned}
\end{equation}
Here 
\begin{equation}
    c = \sqrt{\rho / \chi} \ , \quad u = k_B T / \rho
\end{equation}
are the spin wave velocity and scaled effective temperature, and $\tilde{h} = \rho \, h$.
$\vec{\eta}(x,t)$ and $\vec{\zeta}(x,t)$ represent independent additive Gaussian white noise terms with vanishing means
$\langle \vec{\eta} \rangle = 0 = \langle \vec{\zeta} \rangle$ and the second moments
\begin{widetext}
\begin{equation}
\begin{aligned}
    \left\langle \eta_\alpha(x_1,t_1) \, \eta_\beta(x_2,t_2) \right\rangle &= - 2 \lambda u \, \delta_{\alpha\beta} \,
    \delta(t_1-t_2) \, \nabla^2\delta^{(d)}(x_1-x_2) \ , \\
    \left\langle \zeta_\alpha(x_1,t_1) \, \zeta_\beta(x_2,t_2) \right\rangle &= 2 D u \, \delta_{\alpha\beta} \, \delta(t_1-t_2) 
    \, \delta^{(d)}(x_1-x_2) \ .
\end{aligned}
\end{equation}

The corresponding Janssen--De~Dominicis response functional becomes 
\cite{janssen1976lagrangean,dominicis1976technics} 
\begin{equation}
\begin{aligned}
    \mathcal{A}[\tilde{m}_\alpha,\tilde{n}_\alpha,m_\alpha,n_\alpha] = \int dt \, d^dx \, \Big[ \tilde{m}_\alpha 
    \left( \partial_t - \lambda \nabla^2 \right) m_\alpha + \tilde{n}_\alpha \left( \partial_t - D \nabla^2 \right) n_\alpha 
    + c \, \epsilon_{\alpha\beta\gamma} 
    \left( \tilde{n}_\alpha \, n_\beta \, m_\gamma - \tilde{m}_\alpha \, n_\beta \, \nabla^2 n_\gamma  \right) \\ 
    + \lambda u \, \tilde{m}_\alpha \, \nabla^2 \tilde{m}_\alpha - D u \, \tilde{n}_\alpha \, \tilde{n}_\alpha \Big] \ ,
\end{aligned}
\end{equation}
\end{widetext}
and the probability distribution $P[\vec{m},\vec{n}]$ for any specific configuration $m_\alpha(x,t)$, $n_\alpha(x,t)$ is
\begin{equation}
    P[\vec{m},\vec{n}] \propto \delta\left( \vec{n}^2 - 1 \right) \int \mathcal{D}[i\tilde{m}_\alpha,i\tilde{n}_\alpha] \, 
    e^{- \mathcal{A}[\tilde{m}_\alpha,\tilde{n}_\alpha,m_\alpha,n_\alpha]} \ .
\end{equation}
As a consequence of the stringent local constraint $\vec{n}(x,t)^2 = 1$, only two components of the three-dimensional 
order parameter field $\vec{n} = \left (\vec{\pi},\sigma \right) $ are independent, while the third one is completely 
determined, $\sigma^2 = 1 - \vec{\pi}^2$.
One may thus integrate out the ``longitudinal" $\sigma$ mode and only consider the ``transverse" $\vec{\pi}$ fluctuations.
In order to properly implement the constraint in the ensuing stochastic dynamics, one also needs to impose the noise field 
for the order parameter $\vec{\zeta}$ to be perpendicular to the instantaneous order parameter itself.
In the framework of the associated Janssen--De~Dominicis functional this implies $\tilde{n}_\alpha n_\alpha = 0$ 
\cite{bausch1980nonlinear,fedorenko2006critical}.  

Deep in the ordered phase, the transverse Goldstone mode fluctuations $\vec{\pi}^2$ should be small and one may expand
\begin{equation}
\begin{aligned}
    \sigma &= \sqrt{1-\vec{\pi}^2} = 1 - \frac{\vec{\pi}^2}{2} + \mathcal{O}(\pi^4) \ , \\
    \tilde{\sigma} &= - \tilde{\pi}_i \pi_i / \sqrt{1-\vec{\pi}^2} = 
    - \tilde{\pi}_i \pi_i - \frac{1}{2} \tilde{\pi}_i \pi_i \vec{\pi}^2 + \mathcal{O}(\pi^4)
\end{aligned}
\end{equation}
(where $i=1,2$), which allows the elimination of the $\sigma$ component and the associated $ \tilde{\sigma}$ field.  
The resulting Janssen--De~Dominicis functional becomes 
$\mathcal{A} = \mathcal{A}_\mathrm{h} + \mathcal{A}_{\mathrm{int}}$ with
\begin{widetext}
\begin{equation}
\begin{aligned}
    \mathcal{A}_\mathrm{h} = \int_{t,x} &\Big[ \tilde{m}_i \left( \partial_t - \lambda \nabla^2 \right) m_i + 
    \tilde{m}_3 \left( \partial_t - \Gamma \nabla^2 \right) m_3 - c \, \epsilon_{ij} \tilde{\pi}_i m_j + 
    c \, \epsilon_{ij} \tilde{m}_i \nabla^2 \pi_j - c \tilde{h} \, \epsilon_{ij} \tilde{m}_i \pi_j \\
   &+ \tilde{\pi}_i \left( \partial_t - D \nabla^2 + D \tilde{h} \right) \pi_i - D u \, \tilde{\pi}_i \tilde{\pi}_i + \lambda u \, 
   \tilde{m}_i \nabla^2 \tilde{m}_i + \Gamma u \, \tilde{m}_3 \nabla^2 \tilde{m}_3 + \mathcal{O}(\pi^4) \Big] 
\end{aligned}
\end{equation}
and
\begin{equation}
\begin{aligned}
    \mathcal{A}_{\mathrm{int}} = \int_{t,x} &\Big[ \frac{\tilde{\pi}_i \pi_i}{2} \left( \partial_t - D \nabla^2 + D \tilde{h}
    \right) \vec{\pi}^2 - D u (\tilde{\pi}_i \pi_i)^2 + c \, \epsilon_{ij} \tilde{\pi}_i \pi_j m_3 - 
   c \, \epsilon_{ij} \tilde{m}_3 \pi_i \nabla^2 \pi_j \\
   -c &\, \epsilon_{ij} (\tilde{\pi}_k \pi_k) \pi_i m_j + \frac{c \, \epsilon_{ij}}{2} \, \tilde{\pi}_i m_j \vec{\pi}^2 - 
   \frac{c \, \epsilon_{ij}}{2} \, (\tilde{m}_i \nabla^2\pi_j) \vec{\pi}^2 + 
   \frac{c \, \epsilon_{ij}}{2} \, \tilde{m}_i \pi_j \nabla^2 \vec{\pi}^2 + \mathcal{O}(\pi^4) \Big] ,
\end{aligned}
\end{equation}
where $\int_{t,x} \ldots = \int dt \, d^dx \ldots$ and $i,j,k=1,2$.
\end{widetext}

A cutoff in the $\vec{\pi}^2$ expansion naturally introduces explicit symmetry breaking and manifestly leaves us in the 
ordered phase region, which is governed by a so-called coexistence RG fixed point \cite{lawrie1981goldstone}. 
The explicit symmetry breaking induces anisotropy for the different order parameter components.
In anticipation of different renormalizations for the transverse and longitudinal components of the conserved magnetization, 
we thus introduce a distinct diffusivity $\Gamma$ for the $m_3$ component of the magnetization density. 
Yet this explicit symmetry breaking is expected to be resolved by the RG flow upon approaching the critical point, where 
one should recover $\Gamma = \lambda$. \\

\subsection{Dynamic response functions}

To retrieve the dynamical response functions for our model, we introduce source terms to $\vec{\pi}$ and $\vec{M}$ in the 
Hamiltonian (\ref{Hamiltonian})
\begin{equation}
    \mathcal{H}_s[\vec{M},\vec{\pi}] = - \int d^dx \left( \rho \, J_i \pi_i + c \, M_\alpha H_\alpha \right) ,
\end{equation}
where $i=1,2$ and $\alpha=1,2,3$. 
The thus induced source terms in the Janssen--De~Dominicis functional are
\begin{widetext}

\begin{equation}
\begin{aligned}
    \mathcal{A}_s =& \int_{t,x} \Big[ \lambda \, \tilde{m}_i \nabla^2 H_i + \Gamma \, \tilde{m}_3 \nabla^2 H_3 
    - D \, \tilde{\pi}_i J_i - c \, \tilde{m}_3 \epsilon_{ij} \pi_i J_j + c \, \epsilon_{ij} \tilde{m}_i J_j
    - \frac{c \, \epsilon_{ij}}{2} \tilde{m}_i J_j \vec{\pi}^2 - c \, \epsilon_{ij} \tilde{\pi}_i \pi_j H_3\\ &+ c \, (\tilde{\pi}_k \pi_k) \epsilon_{ij} \pi_i H_j 
    + c \, \epsilon_{ij} \tilde{\pi}_i H_j - \frac{c \, \epsilon_{ij}}{2} \tilde{\pi}_i H_j \vec{\pi}^2
    - c \, \epsilon_{ij} \tilde{m}_i m_j H_3  
    +c \, \epsilon_{ij} \tilde{m}_i H_j m_3 - c \, \tilde{m}_3 \epsilon_{ij} m_i H_j + \mathcal{O}(\pi^4) \Big] \ .
\end{aligned}
\end{equation}
With the usual definition of the dynamical response functions
\begin{equation}
\begin{aligned}
    \chi_{ij}(x,t;x',t') &=\left. \frac{\delta \langle \pi_i(x,t) \rangle_s}{\delta J_j(x',t')} \right|_{H=0,J=0} \, , \\
    R_{ij}(x,t;x',t') &= \left. \frac{\delta \langle m_i(x,t) \rangle_s}{\delta H_j(x',t')} \right|_{H=0,J=0} \, , \quad
    R_3(x,t;x',t') = \left. \frac{\delta \langle m_3(x,t) \rangle_s}{\delta H_3(x',t')} \right|_{H=0,J=0} \, ,
\end{aligned}
\end{equation}
and the probability distribution given by the Janssen--De~Dominicis functional, one may express the response functions in
terms of the correlation functions in the field theory, 
\begin{equation}
\begin{aligned}
   \chi_{ij}(x,t;x',t') &= D \, \langle \pi_i(x,t) \tilde{\pi}_j(x',t') \rangle +  c \, \epsilon_{kj} \langle \pi_i(x,t) 
   [\tilde{m}_3 \pi_k](x',t') \rangle 
   - c \, \epsilon_{kj} \langle\pi_i(x,t) \tilde{m}_k(x',t') \rangle \\ &\quad+ \frac{c}{2} \, \epsilon_{kj} \langle \pi_i(x,t)
  [\tilde{m}_k \vec{\pi}^2](x',t') \rangle + \mathcal{O}(\pi^4) \ , \\
  R_{ij}(x,t;x',t') &= - \lambda \, \langle m_i(x,t) \nabla^2 \tilde{m}_j(x',t') \rangle - c \, \epsilon_{kj} \langle m_i(x,t)
  [\tilde{\pi}_l \pi_l \pi_k](x',t') \rangle  + \frac{c}{2} \, \epsilon_{kj} \langle m_i(x,t)
  [\tilde{\pi}_k \vec{\pi}^2](x',t') \rangle \\
  &\quad - c \, \epsilon_{kj} \langle m_i(x,t) [\tilde{m}_k m_3](x',t') \rangle + c \, \epsilon_{kj} \langle m_i(x,t) 
  [\tilde{m}_3m_k](x',t') \rangle   - c \, \epsilon_{kj} \langle m_i(x,t) \tilde{\pi}_k(x',t') \rangle + \mathcal{O}(\pi^4) \ , \\
  R_\parallel(x,t;x',t') &= - \Gamma \, \langle m_3(x,t)\nabla^2 \tilde{m}_3(x',t') \rangle + c \, \epsilon_{ij} \langle m_3(x,t)
  [\tilde{m}_i m_j](x',t') \rangle + c \, \epsilon_{ij} \langle m_3(x,t) [\tilde{\pi}_i \pi_j](x',t') \rangle + \mathcal{O}(\pi^4)
  \ .
\end{aligned}
\label{susc}
\end{equation}
\end{widetext}
We note that the correlation functions with subscript $s$ are calculated in the presence of external source terms, while the 
correlation functions without subscripts are calculated without external sources.

We have evalulated the dynamical response functions to one-loop order with respect to the transverse fields, which 
corresponds to a first-order perturbation in the effective temperature $u$ for the statics and  the parameter $u \, c^2$ for
dynamical quantities. 
By means of a Dyson equation resummation of the one-loop order results, the dynamical susceptibilities can be written in the 
following compact form \cite{tauberdissertation}, 
\begin{widetext}

\begin{equation}
\begin{aligned}
    \chi_{ij}(p,\omega) &= \delta_{ij} \chi(p,0) \left[ 1 + \frac{i\omega}{-i\omega + \Delta(p,\omega) + C(p,\omega)^2 /
    \left[ -i\omega + \Lambda(p,\omega) \right]} \right] , \\
    R_{ij}(p,\omega) &= \delta_{ij} R(p,0) \left[ 1 + \frac{i\omega}{-i\omega + \Lambda(p,\omega) + C(p,\omega)^2 /
    \left[ -i\omega + \Delta(p,\omega) \right]} \right] , \\
    R_\parallel(p,\omega) &= \frac{\Lambda_\parallel(p,\omega)}{-i\omega + \Lambda_\parallel(p,\omega)} \ ;
\end{aligned}    
\label{response}
\end{equation}
some additional computation details can be found in Appendix A.
Here we have used the static susceptibilities
\begin{equation}
    \chi(p,0) = \frac{1}{p^2+\tilde{h}} \left[ 1 - u \int_k \frac{1}{k^2+\tilde{h}} \right] \, , \quad R(p,0) = 1 \ ,
\end{equation}
with $\int_k \ldots = \int \ldots d^dk / (2 \pi^d) $ and introduced the abbreviations 
\begin{equation}
\begin{aligned}
    C(p,\omega) &= c \, (p^2 + \tilde{h})^{1/2} \left[ 1 - \frac{u}{2} \int_k \frac{1}{k^2 + \tilde{h}} \right] \ , 
    \quad \Lambda(p,\omega) = \lambda p^2 \ , \\
    \Delta(p,\omega) &= D \, (p^2 + \tilde{h}) \ \bigg[ 1 - \frac{c^2 u}{D} \int_k \frac{1}{k^2 + \tilde{h}}  
    \frac{i\omega - \Gamma(p-k)^2 - \lambda k^2}
    {[ -i\omega + i\omega_+(k) + \Gamma(p-k)^2] \, [-i\omega + i\omega_-(k) + \Gamma(p-k)^2]} \bigg] \ , \\
    \Lambda_\parallel(p,\omega) &= \Gamma p^2 - 2 c^2 u \int_k \frac{\left[ k^2 - (p-k)^2 \right] 
    [ i\omega - \lambda k^2 - D (p-k)^2 - D \tilde{h}]}{(k^2 + \tilde{h}) \, [(p-k)^2 - \tilde{h}] 
    \left[ \omega - \omega_+(k) - \omega_+(p-k) \right]} \\
    &\ \times \frac{[ i\omega - \lambda(p-k)^2 - D\tilde{h}] \left[ i\omega - \lambda k^2 - \lambda(p-k)^2 \right] 
    + 2 c^2 (k^2 + \tilde{h})}{\left[ \omega - \omega_-(k) - \omega_+(p-k) \right] 
    \left[ \omega - \omega_-(k) - \omega_-(p-k) \right] \left[ \omega - \omega_+(k) - \omega_-(p-k) \right]} \ , 
\end{aligned}
\label{1-loop results}
\end{equation}
with the Gaussian (mean-field) spin wave dispersion relation
\begin{equation}
    \omega_\pm(p) = - \frac{i}{2} \left[ (D + \lambda) \, p^2 + D \tilde{h} \right] \pm \sqrt{c^2 \, (p^2 + \tilde{h})
    - \frac{1}{4} \left[ (D - \lambda) \, p^2 + D \tilde{h} \right]^2} \ .
\label{dispersion1}
\end{equation}
\end{widetext}
In the limit $\omega \rightarrow 0$, the static response functions are recovered. 
Thus, the dynamics is cleanly separable from the statics.
From the results (\ref{1-loop results}), we also see that in comparison with the SSS model in the ordered phase 
\cite{tauberdissertation}, all contributions from transverse magnetization fluctuations are absent, as a result of the truncated 
$\vec{\pi}^2$ expansion. 

\section{Dimensional $\epsilon$ expansion and scaling properties}

In this section, we will implement the dimensional $\epsilon$ expansion to approach the scaling behavior near RG fixed 
points. 
We start by explaining why a straightforward $d = 2 + \varepsilon$ expansion as appropriate for the static behavior of the
nonlinear sigma model fails to capture its dynamics in the vicinity of the critical fixed point.
We then explore the dynamical scaling behavior at the ordered ``coexistence" phase fixed point with an $\epsilon = 4 - d$ expansion near the model's dynamical upper critical dimension, and subsequently discuss the intriguing crossover features of 
the ensuing RG flows.

\subsection{Failure of the (static) $2 + \varepsilon$ expansion at the critical fixed point}

In the static nonlinear sigma model, it is well-known that perturbatively the transverse fluctuation loops scale according to
$\vec{\pi}^2 \propto u \sim T$ and consequently the $\varepsilon = d - 2$ expansion corresponds to a low-temperature 
expansion about the critical temperature $T_c = 0$ in two dimensions. 
The effective temperature $u$ has scaling dimension $\varepsilon$, which suggests it is marginal near the model's 
{\em lower} critical dimension $d_\textrm{lc} = 2$. 
Therefore, the standard procedure to perturbatively access the critical behavior of the nonlinear sigma model relies on an 
$\varepsilon$ expansion near $d_\textrm{lc} = 2$. 
However, the effective dynamical coupling for the reversible mode-coupling terms is  $c^2u$, as can be inferred from 
Eq.~(\ref{1-loop results}). 
Its naive scaling dimension is $[c^2u] = 4 - d$, resulting in a dynamical {\em upper} critical dimension $d_c = 4$.  

At dimension $d = 2 + \varepsilon$, we have $[c^2u] = 2 - \varepsilon$, which is relevant in the infrared scaling regime,
and hence will tend to infinity under RG scale transformations near $d_{\textrm lc} = 2$. 
If instead we start near the upper critical dimension $d_c = 4$, setting $d = 4 - \epsilon$, the scaling dimension of the 
static expansion parameter $u$ becomes $\epsilon - 2$, which is irrelevant in the infrared.
Yet in the static nonlinear sigma model, information about the asymptotic infrared scaling behavior is extracted from an
ultraviolet-stable RG fixed point that controls the short-distance divergences for $d > d_\textrm{lc} = 2$. 
For the static ultraviolet fixed point, near dimension $d_c = 4$, however, the parameter $u$ becomes relevant. 
Thus, a standard, straightforward truncation over $\vec{\pi}^2 \propto u$ is improper as the corresponding perturbative 
expansion diverges near $d_c = 4$, and cannot be controlled at that static fixed point. 
In order to properly retrieve the scaling information near the ultraviolet RG fixed point near dimension $d_c = 4$, we would 
need to apply a resummation of all diagrams related to the $\vec{\pi}^2$ expansion, which appears to be unfeasible with
currently available tools.

In conclusion, in stark contrast with its counterpart with purely relaxational kinetics, the dynamical critical behavior of the 
nonlinear sigma model with reversible mode-coupling terms cannot be simultaneously analyzed along with the static critical 
properties by means of a perturbative dimensional expansion either near $d_\textrm{lc} = 2$ or $d_c = 4$.

\subsection{Scaling behavior near the ordered phase coexistence fixed point}

In the low-temperature phase with spontaneously broken rotational symmetry $O(n)$, the vectorial order parameter 
separates into an ``Anderson--Higgs" mode $\sigma$ and $n-1$ massless Goldstone modes $\vec \pi$. 
We note, though, that the Higgs mode appears massive only in the mean-field (Gaussian) approximation, and below 
$d_c = 4$  is rendered massless too as a consequence of its coupling to the strong transverse fluctuations \cite{anderson1958random,brezin1973critical,nelson1976coexistence,lawrie1981goldstone,tauber1992critical}. 
In an infinite system, the massless Goldstone modes induce universal static and dynamic scaling behavior that is distinct from
the critical scaling properties.
This coexistence limit is captured in the RG treatment by an infrared-stable zero-temperature coexistence fixed point.
We thus anticipate novel universal scaling behavior for the massless Goldstone modes as well as the conserved magnetization density component that is dynamically coupled to the transverse order parameter fluctuations. 
In the following, we study the scaling behavior of the antiferromagnetic nonlinear sigma model with reversible mode 
couplings near this ordered phase fixed point.

To reach the coexistence fixed point, we need to tune the effective temperature $u \to 0$ and work in a dimensional 
$\epsilon = 4 - d$ expansion near the upper critical dimension $d_c = 4$. 
However, since $c^2$ also scales as we approach the fixed point and the effective dynamical mode coupling parameter 
$g^2 = c^2 u$ becomes marginal near four dimensions, we need to keep $g^2$ fixed when we set $u=0$.
In this low-temperature coexistence regime, we define the multiplicative renormalization $Z$ factors that are to absorb the 
ensuing ultraviolet divergences as follows,
\begin{equation}
\begin{aligned}
    D_R &= Z_D D \, ,\ \ \ \lambda_R  = Z_\lambda \lambda \, , \ \ \ \Gamma_R = Z_\Gamma \Gamma \, , \\ 
    g^2_R &= Z_g \, g^2 A_d \, \mu^{-\epsilon} \, , \ \ \  c^2_R = Z_c \, c^2 \mu^{-2} \, ,
\end{aligned}
\end{equation}
where $\mu$ labels an arbitrary momentum scale at which this renormalization is carried out, and 
$A_d = \Gamma(3-d/2) / 2^{d-1} \pi^{d/2}$ is a geometric factor.
Our choice of the normalization point, which only must reside outside the infrared-singular region, lies at $\tilde{h} = 0$,
$p = 0$, and $C(p,\omega) \sqrt{R(p,0) \chi(p,0)} = c \mu$ \cite{tauberdissertation}. 

To one-loop order, the $Z$ factors can then be inferred from the explicit expressions (\ref{1-loop results}), which yields
\begin{equation}
\begin{aligned}
    Z_\lambda &= Z_g = Z_c = 1 \ , \\
    Z_\Gamma &= 1 + \frac{f}{2 (1+w)} \frac{A_d \mu^{-\epsilon}}{\epsilon} 
    \left[ w + \left( 1 + \frac{c^2}{D \lambda \mu^2} \right)^{-\epsilon/2} \right] \ , \\
    Z_D &= 1 + \frac{f}{1+v} \frac{A_d \mu^{-\epsilon}}{\epsilon} 
    \left[ 1 + \frac{c^2}{(\Gamma + D) (\Gamma + \lambda) \mu^2} \right]^{-\epsilon/2} \ ,
\end{aligned}
\end{equation}
where for convenience we have defined the dimensionless ratios of relaxation times $w, v$ and the effective mode coupling strength $f$ as
\begin{equation}
    w = D / \lambda \ , \quad v = D / \Gamma \ , \quad f = g^2 / D \Gamma \ .
\end{equation}
The scale-dependent anomalous dimensions (Wilson's RG flow functions) thus follow  
\begin{equation}
\begin{aligned}
    \zeta_\lambda &= \left. \mu \frac{\partial}{\partial \mu} \right|_0 \ln \frac{\lambda_R}{\lambda} =
    \left. \mu \frac{\partial}{\partial \mu} \right|_0 \ln Z_\lambda = 0 \ , \\
    \zeta_g &= \left. \mu \frac{\partial}{\partial \mu} \right|_0 \ln \frac{g^2_R}{g^2} = 
    - \epsilon + \left. \mu \frac{\partial}{\partial \mu} \right|_0 \ln Z_g = - \epsilon \ , \\
    \zeta_\Gamma &= \left. \mu \frac{\partial}{\partial \mu} \right|_0 \ln \frac{\Gamma_R}{\Gamma} = 
    \left. \mu \frac{\partial}{\partial \mu} \right|_0 \ln Z_\Gamma \\
    &= - \frac{f_R}{2 (1+w_R)}  \left[ w_R + \left( 1 + \frac{c_R^2}{D_R \lambda_R} \right)^{-(1 + \epsilon/2)} 
    \right] , \\
    \zeta_D &= \left. \mu \frac{\partial}{\partial \mu} \right|_0 \ln \frac{D_R}{D} = 
    \left. \mu \frac{\partial}{\partial \mu} \right|_0 \ln Z_D \\
    &= - \frac{f_R}{1+v_R} \left[ 1 + \frac{c_R^2}{(\Gamma_R + D_R) (\Gamma_R + \lambda_R)} 
    \right]^{-(1 + \epsilon/2)} . 
\end{aligned}
\label{zetaf}
\end{equation}

These anomalous dimensions enter the fundamental RG equations for the correlation and vertex functions as well as dynamical
susceptibilites, which capture the behavior of the model under the change of renormalization scales \cite{cardy1996scaling,zinn2021quantum,tauber2014critical}.
One may then invoke the method of characteristics to solve the RG partial differential equations. 
By introducing the flow parameter $\mu(l) = \mu l$, the RG equations can be separated into a set of coupled first-order 
differential flow equations. 
Though we seem to require many parameters, they are not completely independent from each other, and in the end we 
merely need the following four independent flow equations to describe the entire RG flow of the system,
\begin{equation}
\begin{aligned}
    l \frac{\mathrm{d} f(l)}{\mathrm{d} l} &= \beta_f(l) = f(l) \, \Big[ \zeta_g(l) - \zeta_D(l) - \zeta_\Gamma(l) \Bigr] \ , \\
    \quad l \frac{\mathrm{d} c^2(l)}{\mathrm{d} l} &= -2 \, c^2(l) \ , \\
    l \frac{\mathrm{d} D(l)}{\mathrm{d} l} &= D(l) \, \zeta_D(l) \ , \quad
    l \frac{\mathrm{d} \Gamma(l)}{\mathrm{d} l} = \Gamma(l) \, \zeta_\Gamma(l) \ ,
\end{aligned}
\label{floweqs}
\end{equation}
where the dependence of the RG zeta functions on the flow parameter $l$ stems from the $l$-dependence of all the 
parameters on the r.h.s. of Eqs.~(\ref{zetaf}).
In terms of the anomalous dimensions, the scale-dependent effective dynamical exponents can be defined via
\begin{equation}
     z_\lambda(l) = 2 \, , \ z_\Gamma(l) = 2 + \zeta_\Gamma(l) \, , \ z_D(l) = 2 + \zeta_D(l) \ . 
\end{equation}
Since the transverse magnetization diffusivity $\lambda$ acquires no fluctuation correction at one-loop order, the dynamic 
exponent for the perpendicular direction of the magnetization density retains its Gaussian value $z_\lambda = 2$, indicating 
standard diffusive relaxation.

The flow equation for the renormalized spin wave speed is immediately solved by $c(l) = c(1) / l$ with $c(1) = c_R$.
Consequently, three distinct fixed points affect the RG flows:
(i) The Gaussian fixed point $f_G^* = 0$ with mean-field dynamic scaling exponents $z_D = z_\lambda = z_\Gamma = 2$ 
is infrared-unstable (in the $f$ direction) below the upper critical dimension $d_c = 4$, but governs the ultraviolet region for 
large $l > 100$.

(ii) Setting $c_R = 0$, we obtain the (quasi-)critical fixed point, albeit with the Anderson--Higgs mode ``frozen": 
In contrast with the Landau--Ginzburg--Wilson Hamiltonian, the nonlinear sigma model does not permit independent 
longitudinal fluctuations in the $\sigma$ field.
Eqs.~(\ref{zetaf}) reduce to $\zeta_\Gamma = - f_R / 2$, $\zeta_D = - f_R / (1 + v_R)$, and the existence of a nontrivial, 
finite mode coupling fixed point $0 < f^* < \infty$ requires the square bracket in the RG beta function $\beta_f$ to vanish, 
which implies $- \zeta_\Gamma^* - \zeta_D^* = -\zeta_g^* = \epsilon$.
The dimensionless relaxation time scale ratio $v$ is governed by the beta function $\beta_v = v (\zeta_D - \zeta_\Gamma)$,
implying for $v^* > 0$ strong dynamic scaling with $\zeta_D ^* = \zeta_\Gamma^* = - \epsilon / 2$.
This leads to the one-loop fixed point values $v_c^* = 1$ and $f_c^* = \epsilon$, and $z_c = 2 - \epsilon / 2 = d / 2$.
Note that these correspond to the critical fixed point of the $n$-component SSS model (with strong dynamic scaling) 
$v_c^* = 2 n - 3$ and $f_c^* = \epsilon$ \cite{de1978field,folk2006critical,tauber2014critical} for $n = 2$, i.e., effectively 
the planar model E \cite{halperin1976renormalization}, {\em not} the three-component model G applicable to the critical 
dynamics of isotropic Heisenberg antiferromagnets 
\cite{schwabl1970hydrodynamics,halperin1974renormalization,halperin1976renormalization}. 
This reflects the suppressed longitudinal $\sigma$ fluctuations; but since the dynamical critical exponent is in fact 
independent of $n$, the system appears to recover the correct critical scaling properties.

(iii) Finally for $c(1) = c_R > 0$,  the RG flows asymptotically reach the coexistence fixed point as $l \to 0$, 
which govers the universal scaling behavior in the ordered phase at long wavelengths and low frequencies.
With $c(l) = c(1) / l \to \infty$, we have $\zeta_\Gamma = - f_R w_R / [2 (1 + w_R)]$ and $\zeta_D = 0$.
Consequently $v_o^* = 0$, whereas $w_R$ assumes a non-universal value that is determined by the initial conditions of the
RG flow.
A nontrivial RG fixed point value $0 < f_o^* < \infty$ now requires $- \zeta_\Gamma^* = \epsilon$, whence the modified 
effective mode-coupling strength ${\tilde f}_R = f_R w_R / (1 + w_R)$ must assume the universal fixed point value 
${\tilde f}^* = 2 \epsilon$.
The original mode coupling strength $f_R$ however approaches non-universal limiting values as $l \to 0$.
For the associated effective dynamic scaling exponents, we obtain in the coexistence regime
\begin{equation}
     z_D = 2 \ , \quad z_\lambda = 2 \ , \quad z_\Gamma = 2 - \epsilon = d - 2 \ .
\end{equation}
We remark that at the upper critical dimension $d_c = 4$, the longitudinal magnetization density would acquire mere 
logarithmic corrections to its mean-field diffusive relaxation.
Near four dimensions, therefore, the crossover to the asymptotic scaling exponent becomes quite slow, as demonstrated in
Fig.~\ref{z3.9} in Appendix~C.
\begin{figure}[t]
\centering
\subfloat[]{\includegraphics[scale=0.5]{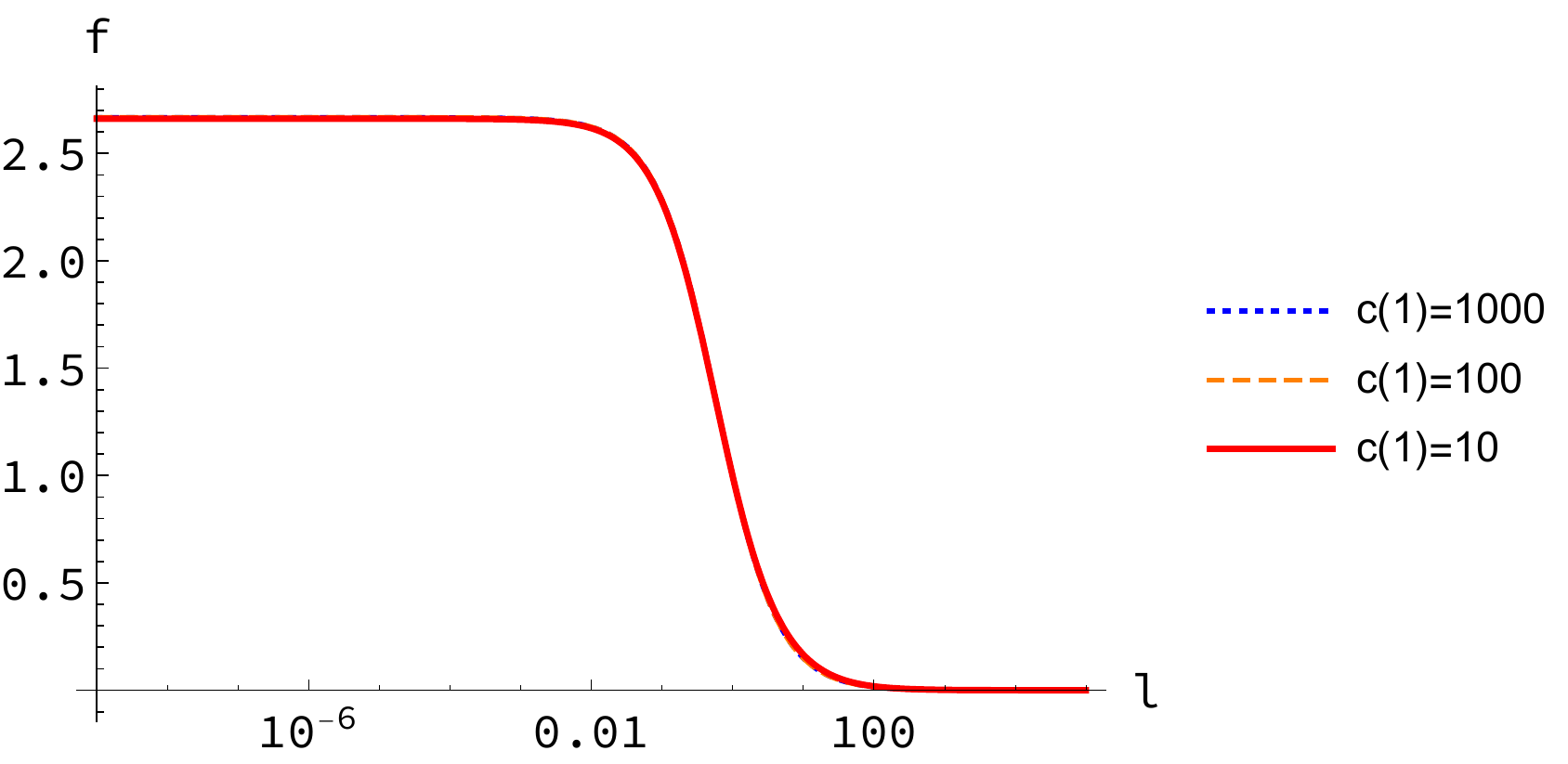}}\\ 
\centering
\subfloat[]{\includegraphics[scale=0.5]{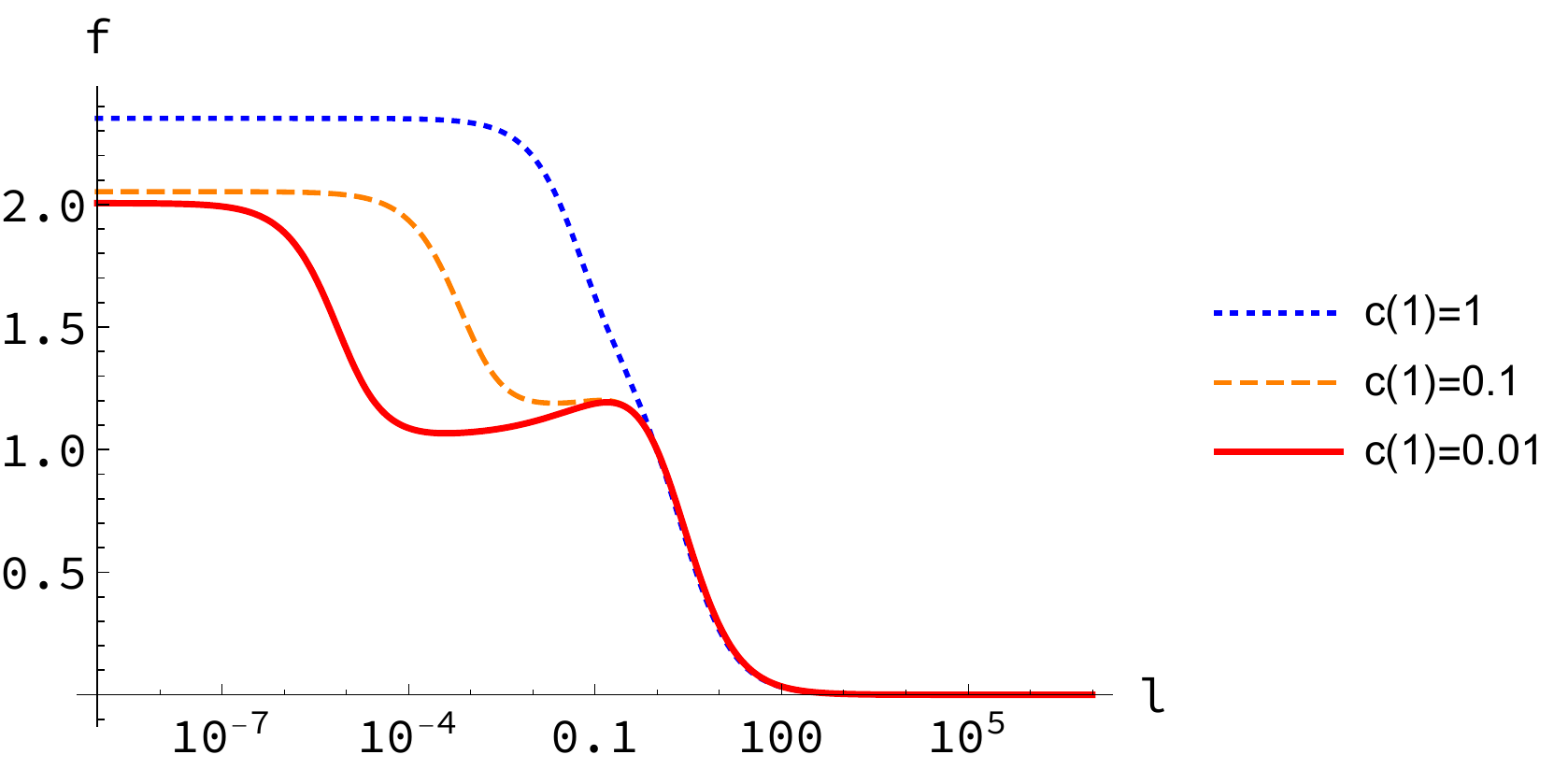}}\\
\centering
\subfloat[]{\includegraphics[scale=0.5]{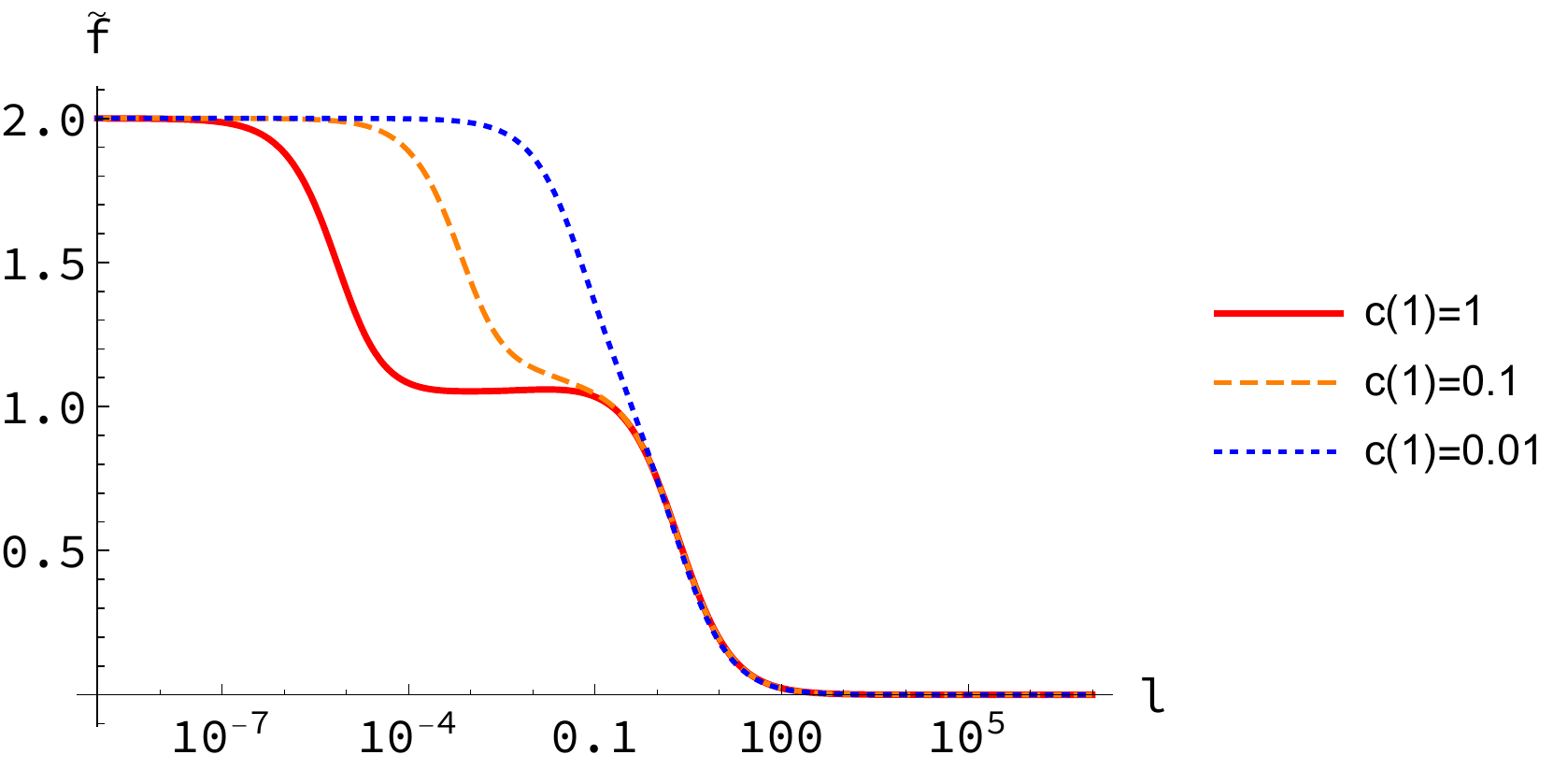}}
\caption{RG flow of the effective mode-coupling strength $f(l)$ with initial values $f(1) = 1$, $D(1) = 3$,
	$\Gamma(1) = 1$, $\lambda(1) = \lambda_R = 1$, and different initial values for $c(1)$ as indicated, in $d = 3$
       dimensions, i.e., for $\epsilon = 1$.
       For large $c(1) \gg 1$ (a), one observes a direct crossover from the unstable Gaussian to the stable coexistence fixed 
       point as $l \to 0$ (all three curves essentially coincide).
       For small $c(1) \leq 1$ (b), the RG flow first runs into the critical fixed point $f_c^* = \epsilon$, but eventually too 
       approaches the asymptotic coexistence regime describing the scaling properties in the ordered phase.
       Whereas $f$ approaches a non-universal value in the coexistence limit, the effective modified mode coupling 
       $\tilde f = w f / (1 + w) \to 2 \epsilon$ universally for arbitrary initial flow conditions (c).}
\label{f}
\end{figure}

\begin{figure}[t]
\centering
\subfloat[]{\includegraphics[scale=0.5]{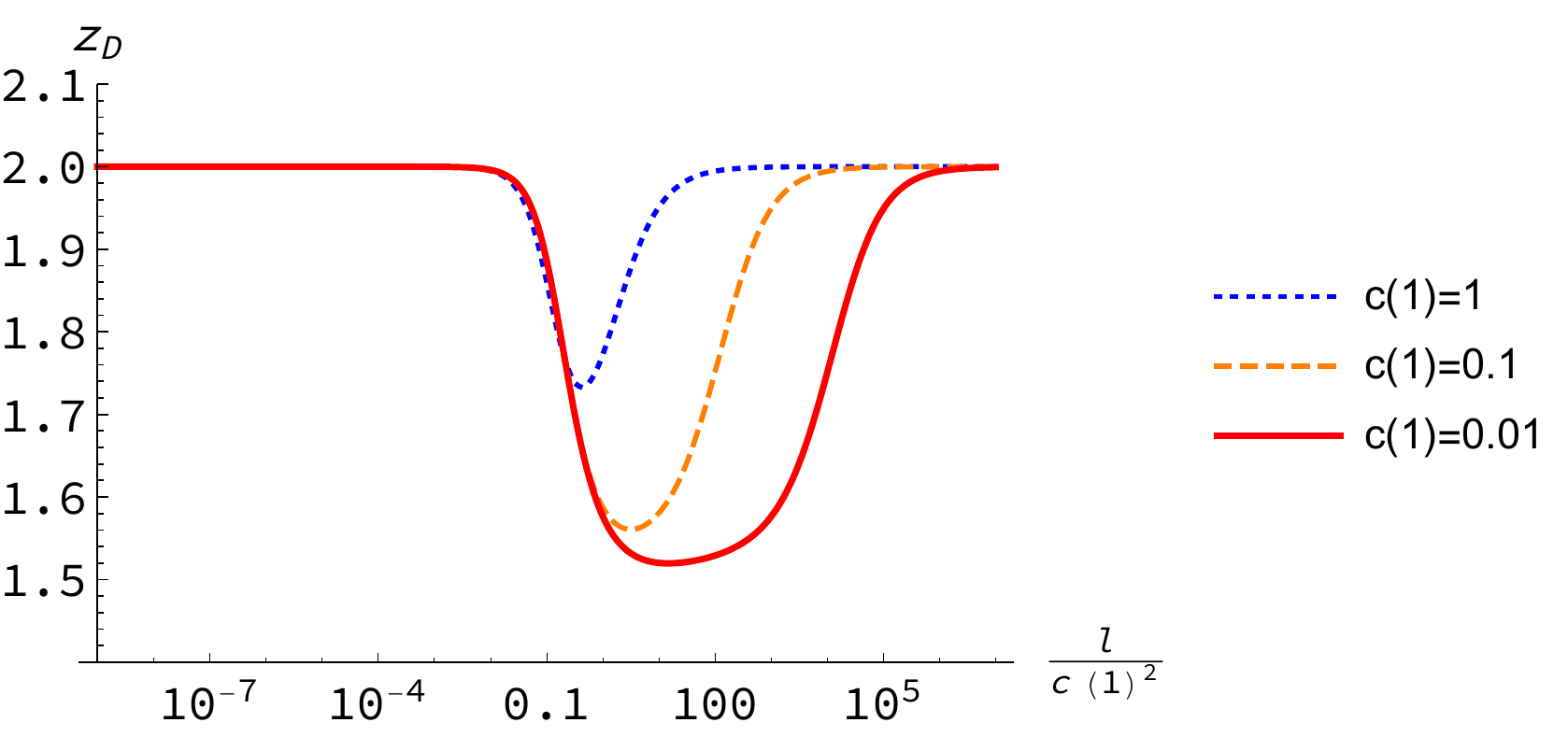}}\\ 
\centering
\subfloat[]{\includegraphics[scale=0.5]{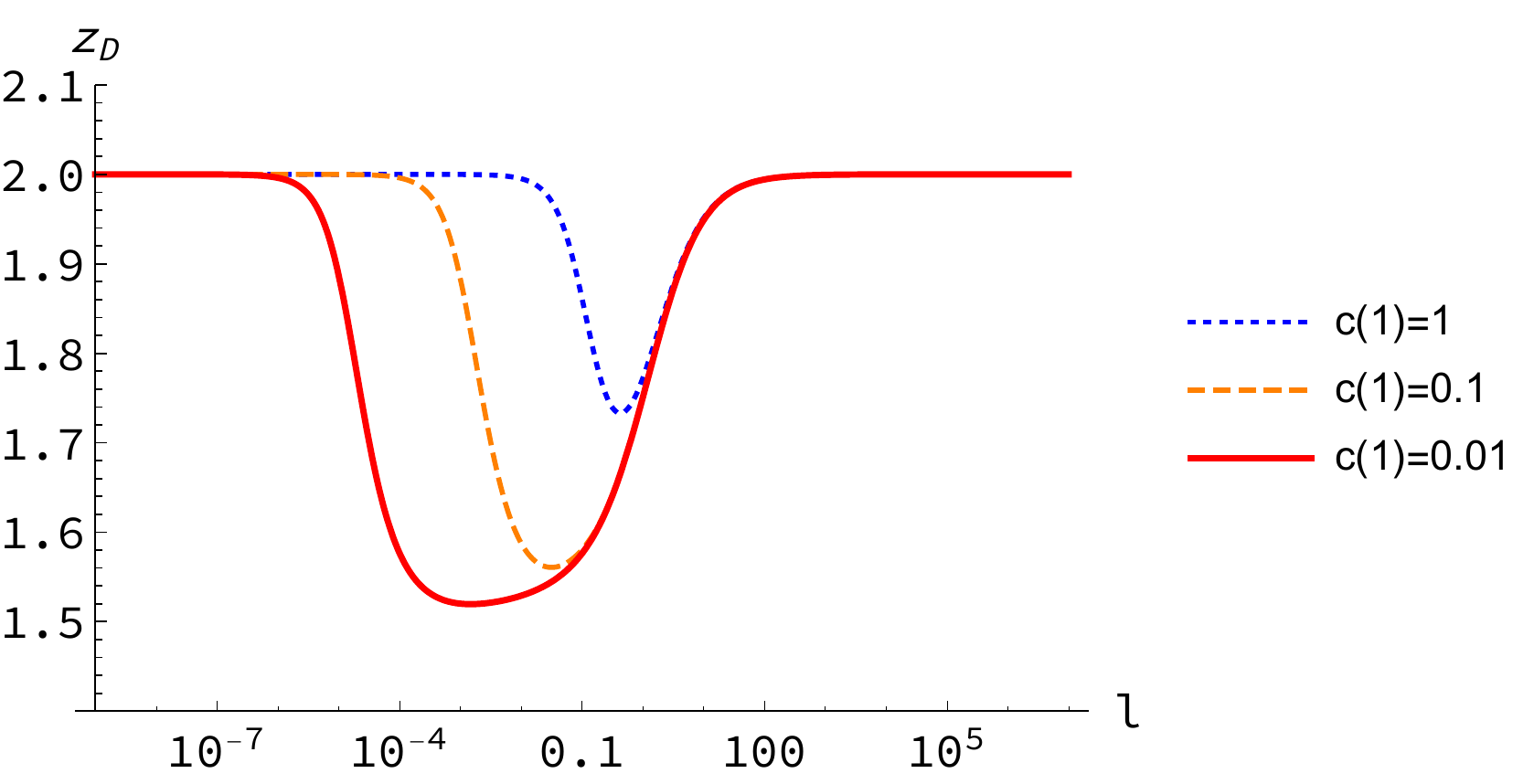}}
\caption{RG flow of the effective dynamical exponent $z_D(l)$ for the relaxation of the transverse order parameter 
	components $\vec \pi$, with initial conditions $f(1) = 1$, $D(1) = 3$, $\Gamma(1) = 1$, 
	$\lambda(1) = \lambda_R = 1$, and various (small) values of $c(1) = 1, 0.1, 0.01$, for $\epsilon = 1$ ($d = 3$). 
 	In (a), the flow parameter $l$ has been rescaled by the initial value $c(1)^2$ to collapse the curves near their approach,
	as $l \to 0$, to the coexistence RG fixed point that governs the low-temperature ordered phase.
      In (b), in contrast the graphs for the same data collapse near the initial Gaussian RG fixed point.
      Both at the Gaussian and coexistence fixed points, $z_D = 2$; whereas at the critical RG fixed point, $z_D = d / 2$.}
\label{zD}
\end{figure}

\begin{figure}[t]
\centering
\subfloat[]{\includegraphics[scale=0.5]{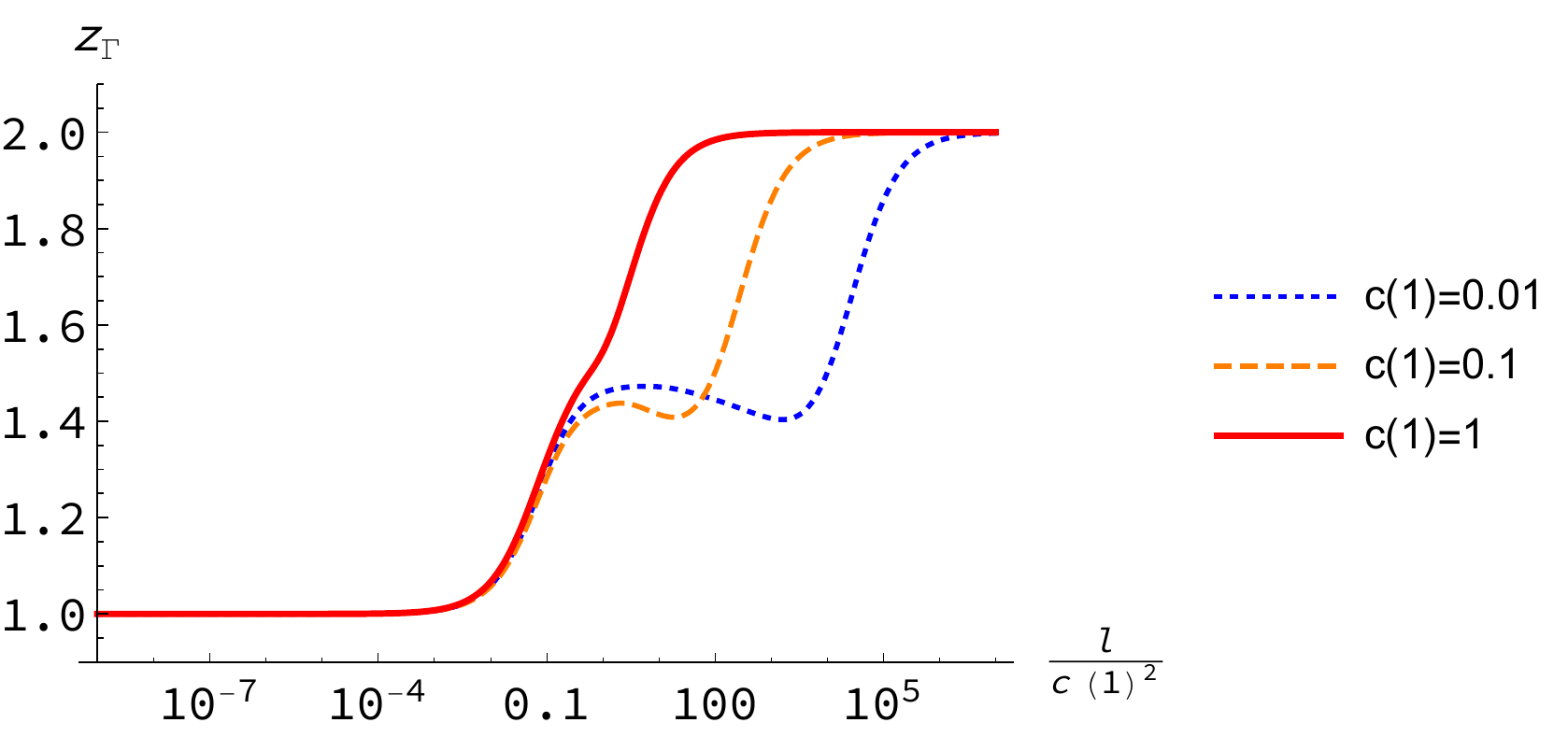}}\\ 
\centering
\subfloat[]{\includegraphics[scale=0.5]{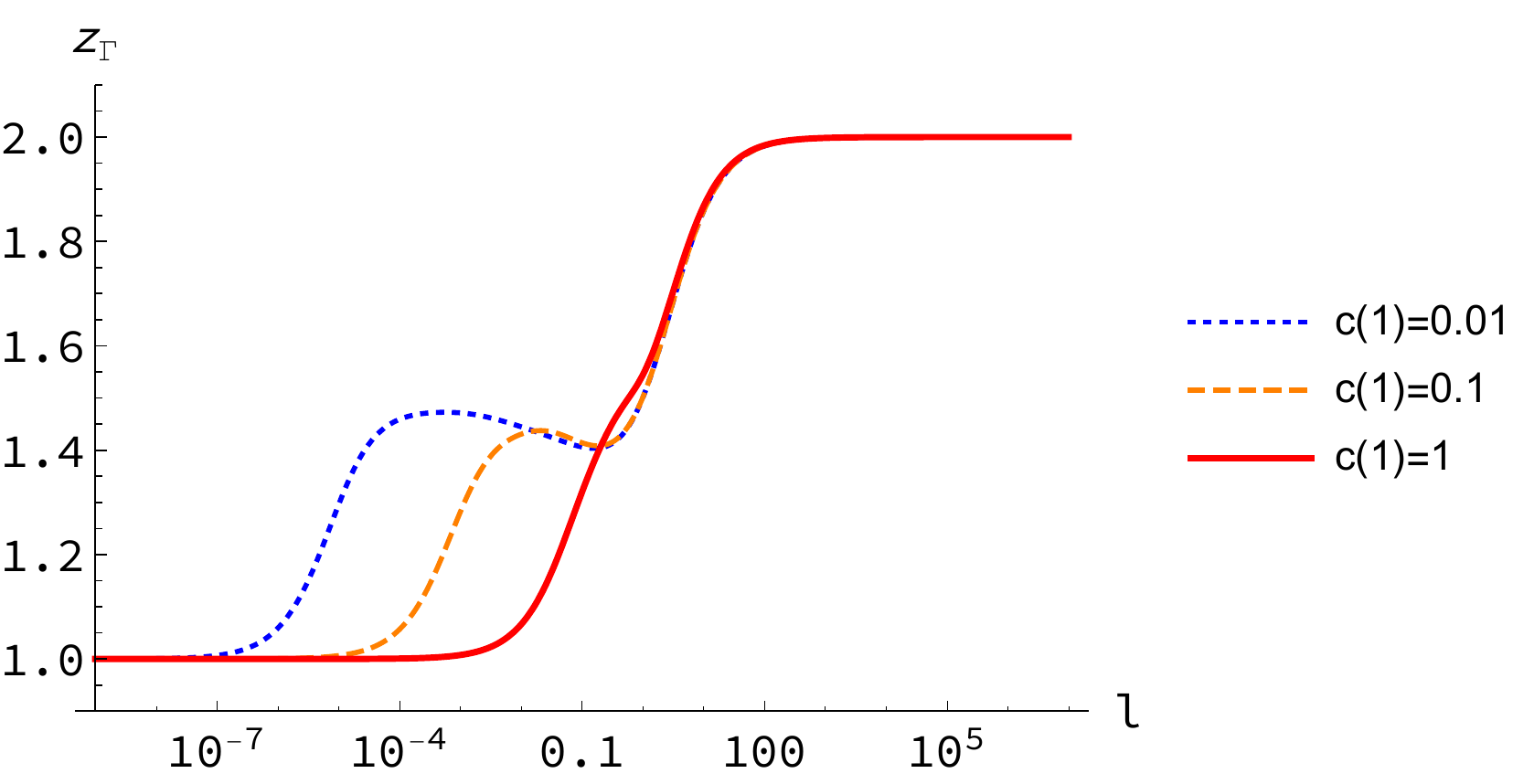}}
\caption{RG flow of the effective dynamical exponent $z_\Gamma(l)$ of the longitudinal component of the magnetization 
	density $m_3$ in $d = 3$ dimensions, with initial values $f(1) = 1$, $D(1) = 3$, $\Gamma(1) = 1$, 
      $\lambda(1) = \lambda_R = 1$, and $c(1)$ as indicated, plotted vs. the flow parameter $l / c(1)^2$ in (a), and vs. $l$ 
      in (b) to demonstrate data collapse in the vicinity of the coexistence and Gaussian fixed points, respectively.
	At the critical RG fixed point, $z_\Gamma = d / 2$; in the ordered phase ($l \to 0$), the longitudinal magnetization 
	relaxation acquires anomalous sub-diffusive scaling, described by $z_\Gamma = 2 - \epsilon = d - 2$.}
\label{zGamma}
\end{figure}

\begin{figure}[t]
\centering
\subfloat[]{\includegraphics[scale=0.34]{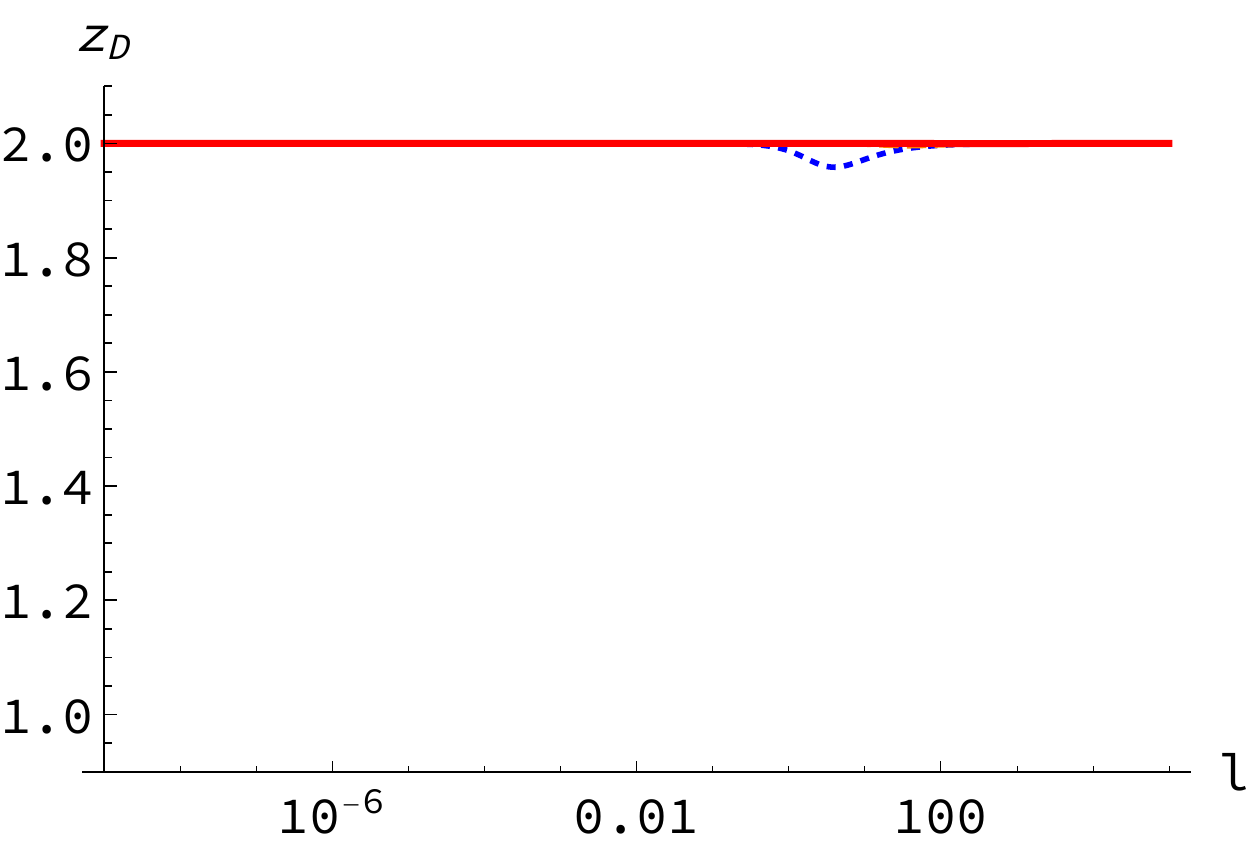}}
\subfloat[]{\includegraphics[scale=0.34]{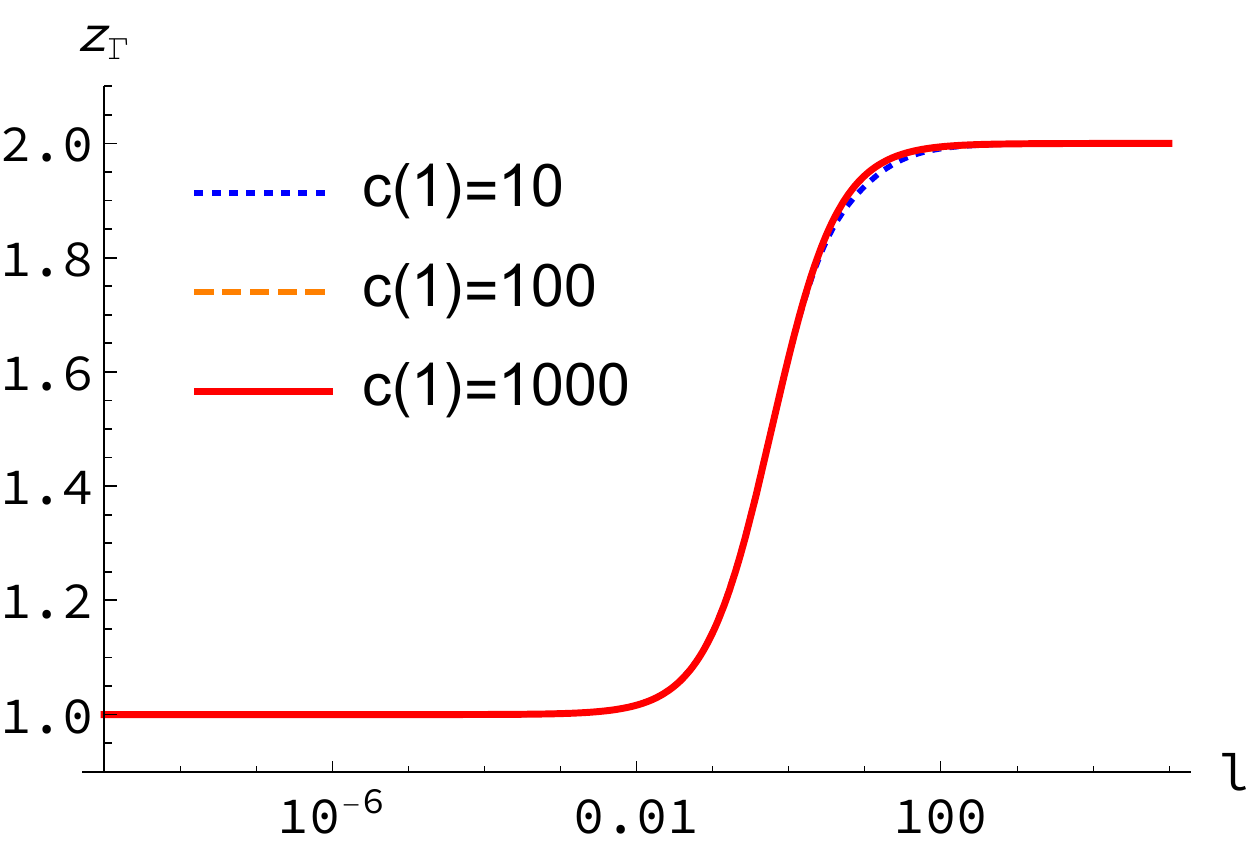}}
\caption{RG flow of the dynamic exponents $z_D(l)$ and $z_\Gamma(l)$ with initial conditions $f(1) = 1$, $D(1) = 3$,
	$\Gamma(1) = 1$, $\lambda(1) = \lambda_R = 1$, and large initial values $c(1) = 10, 100, 1000$.
       The RG flow is then set far away from the critical point, and bypasses the critical scaling regime entirely.}
\label{large c}
\end{figure}

For the relaxation coefficients $D(l)$, $\Gamma(l)$ and the associated effective dynamical exponents, we numerically solve 
the coupled set of flow equations (\ref{floweqs}) in three dimensions ($d = 3$, i.e., $\epsilon = 1$).
(In Appendix C, for illustration we depict the RG flow at $d = 3.9$ or $\epsilon = 0.1$, where the perturbation expansion is more assuredly applicable.)
The ensuing RG flow for the mode-coupling strength $f$ with different initial values for $c(1)$ is shown in Fig.~\ref{f}.
Note that while the initial value was set to $f(1) = 1$, the RG flow was run both towards the ultraviolet regime (large $l$) 
as well as the here relevant infrared region $l \to 0$.
We plot the resulting flows of the effective dynamic exponents $z_D(l)$ and $z_\Gamma(l)$ with different initial conditions in
Figs.~\ref{zD}, \ref{zGamma}, and \ref{large c}. 
The RG flows with small initial values $c(1)$ depicted in Figs.~\ref{f}, \ref{zD}, and \ref{zGamma} indeed approach this 
(quasi-)critical fixed point at intermediate scales, but ultimately depart from it, since $c(l) = c(1) / l$ becomes large in the 
infrared regime $l < 1$.

The explicit numerical solutions of the coupled RG flow equations shown in the figures for $d = 3$ ($\epsilon = 1)$ confirm
the above stability and scaling analysis near the three distinct RG fixed points.
For $l \to 0$, both scale-dependent effective dynamical exponents $z_D(l)$ (Fig.~\ref{zD}) and $z_\Gamma(l)$ 
(Fig.~\ref{zGamma}) flow to their ordered phase fixed point values $z_D = 2$ and $z_\Gamma = 1$. 
Thus, both dynamical exponents of the transverse components of the order parameter and magnetization density are 
$z_D = z_\lambda = 2$ in the coexistence limit, indicating purely diffusive behavior.
In contrast, the dynamical scaling exponent of the longitudinal magnetization density $z_\Gamma$ acquires nontrivial 
fluctuation corrections and anomalous, sub-diffusive relaxation: 
$m_3$ decouples from the perpendicular components of the magnetization density, but its coupling to the transverse order 
parameter fluctuations, i.e., the massless Goldstone modes, yields $z_\Gamma = d - 2$, implying linear scaling in three 
dimensions. 
These findings are consistent with the corresponding analysis of the SSS model in the ordered phase 
\cite{tauberdissertation}. 
In addition, for small intial $c(1)$ values, the RG flows remarkably recover the correct critical dynamical exponent.
Although we know that we cannot truly capture the critical regime of the nonlinear sigma model within the $\epsilon = 4 - d$ 
expansion about the dynamical upper critical dimension, in both Figs.~\ref{zD} and \ref{zGamma} we observe a brief
crossover region towards the critical value $z_D = z_\Gamma = d / 2$ for $c(1) < 1$. 
For RG flows initialized with larger $c(1) > 1$, Fig.~\ref{large c} shows that the crossover to the (quasi-)critical regime 
diminishes gradually and finally disappears. 
For sufficiently large $c(1)$, the RG flow directly connects the ultraviolet-stable Gaussian fixed point to the infrared-stable
ordered phase coexistence fixed point.

Even though the nonlinear sigma model and the SSS model are in the same universality class and acquire the same 
anomalous dynamical scaling behavior in the coexistence limit, we emphasize again that the underlying mechanisms differ. 
In the SSS model, the mass of the Anderson--Higgs mode $\sigma$ flows to infinity under the RG transformations, and the 
longitudinal order parameter fluctuations hence ultimately become suppressed. 
In the nonlinear sigma model, in contrast the magnitude of the order parameter is rigidly fixed. 
Thus, we can only consider the Goldstone mode $\vec{\pi}$ fluctuations, whereas longitudinal fluctuations are determined 
by the constraint. 
Near the ordered phase fixed point, the system is almost frozen and the nonlinear sigma model shows identical asymptotic 
scaling properties as the SSS model. 
Yet away from the coexistence fixed point, the RG flows for the various dynamical parameters of these two different models 
can be quite different.

\begin{figure}
\centering
\includegraphics[scale=0.5]{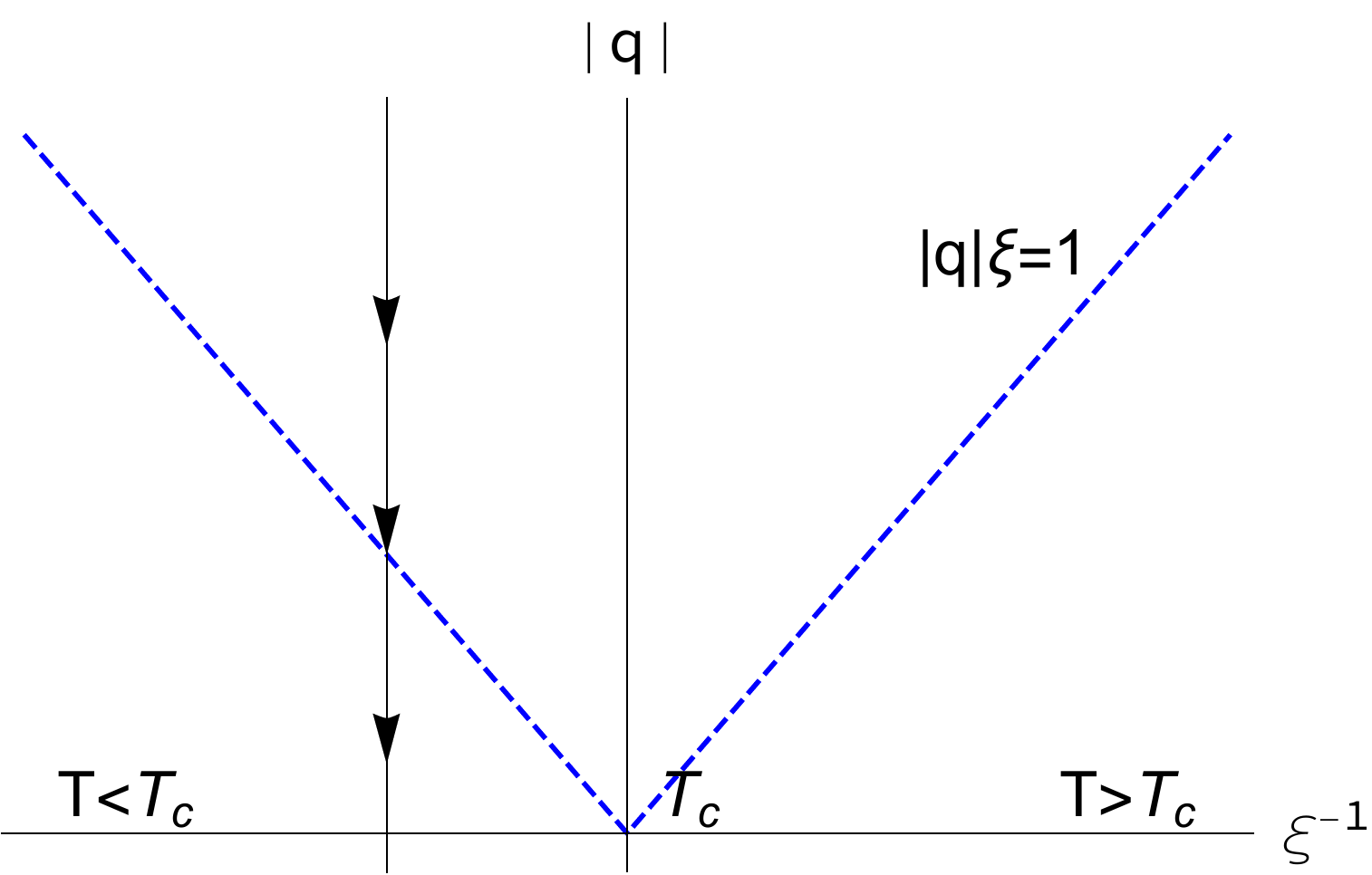}
\caption{Schematic diagram of one RG flow in phase space, corresponding to a wavevector $|q|$ scan in a
    (neutron) scattering experiment at fixed temperature $T < T_c$.}
\label{exp}
\end{figure}

We conclude this discussion with a schematic diagram of one possible scan across $|q|$ in the temperature--wavenumber 
space in Fig.~\ref{exp}, say in an appropriate (neutron) scattering experiment, which actually corresponds to the RG flow 
explored above. 
If we follow the vertical line in the figure, starting in the ordered region for $T<T_c$ with large momentum $|q|$ and subsequently decreasing the wavenumber, we first observe a crossover behavior from the short-distance Gaussian fixed point 
to critical scaling behavior in the region where $|q| \xi \approx 1$. 
Upon further decreasing $|q|$, one reaches the ordered phase coexistence scaling region and ultimately detects the crossover 
from the critical regime to the asymptotic coexistence fixed point.
Deep in the antiferromagnetically ordered phase, and at long wavelengths $q \to 0$, after traversing the crossover region 
with nontrivial scaling behavior, the spin waves originating from the coupled transverse order parameter fields 
(Goldstone modes) and transverse magnetization components will be described asymptotically by the mean-field dispersion 
relation (\ref{dispersion1}) with ballistic propagation $\sim c |q|$ and quadratic damping $D(q) \sim \lambda(q) \sim q^2$, 
albeit with renormalized amplitudes $c, D, \lambda \to c_R, D_R, \lambda_R$.

However, its nonlinear dynamical coupling to the Goldstone modes induces anomalous scaling properties for the longitudinal 
component $m_3$ of the conserved magnetization density field, namely the sub-diffusive wavevector dependence of the 
associated damping coefficient $\Gamma(q) \sim |q|^{z_\Gamma}= |q|^{d-2}$ in $d$ spatial dimensions, i.e., 
$\Gamma(q) \sim |q|$ for $d = 3$, that replaces the corresponding diffusive mean-field behavior $\sim \Gamma q^2$.
In a Lorentzian approximation for the associated dynamical susceptibility $R_\parallel(q,\omega)$ in (\ref{susc}),
$\Lambda_\parallel(q,\omega) \approx \Gamma(q)$, this nontrivial wavevector scaling directly determines the linewidth of 
the longitudinal magnetization correlation function 
$C_\parallel(q,\omega) = (2 k_{\rm B}T / i \omega) \, \textrm{Im}\,R_\parallel(q,\omega)$.
This quantity can be measured experimentally through polarized neutron scattering, although extracting the ultimate 
long-wavelength behavior may be challenging.
Alternatively, spin-echo measurement techniques could provide access to the anomalous temporal decay of the longitudinal 
magnetization autocorrelations $C_\parallel(x=0,t) \sim t^{-d / z_\Gamma} = t^{- d / (d-2)} \sim t^{-3}$ in three 
dimensions, rather than the mean-field decay $C_\parallel(0,t)\sim t^{-d / 2} = t^{-3/2}$ for $d = 3$.

\section{Summary and conclusion}

In this paper, we have investigated the critical dynamics of the antiferromagnetic nonlinear sigma model with conserved total 
magnetization.
The Langevin dynamics then involves reversible, hydrodynamic mode-coupling terms.
We have investigated the ensuing scaling properties by means of the perturbative field-theoretic RG approach, and computed
dynamical response functions of both the order parameter and the magnetization densities to one-loop order.
As expected near thermal equilibrium, the response functions show a separation of the dynamics from the statics.
There emerges an explicit symmetry breaking between the response functions of the transverse and longitudinal components 
of the magnetization density, which suggests that the critical fixed point is no longer approachable in this dynamical nonlinear
sigma model variant, since the rotational symmetry cannot be recovered. 
We have provided a detailed argument from the RG point of view why approaching the critical fixed point is not feasible, at
least in this perturbative regime.

We have analyzed the dynamical scaling behavior near the ordered phase RG fixed point that describes the asymptotic 
coexistence limit via an $\epsilon = 4 - d$ expansion about the dynamical upper critical dimension $d_c = 4$. 
We find Goldstone-mode induced anomalous sub-diffusive scaling for the longitudinal component of the magnetization.
The dynamic exponents at the coexistence ordered phase fixed point coincide with those for the SSS model which is based on
a Landau--Ginzburg--Wilson Hamiltonian.
We have numerically solved the RG flow equations, and observe that if initiated near the critical point, they display a brief crossover regime towards the critical region, even somewhat fortuituously recovering the correct dynamical critical exponent
$z_c = d / 2$, before ultimately reaching the coexistence limit with $z_D = z_\lambda = 2$ and $z_\Gamma = d - 2$. 
These relaxation predictions should be experimentally accessible through the dynamic scaling of neutron scattering linewidths
or the temporal decay of the autocorrelation function for the longitudinal magnetization fluctuations.

\begin{acknowledgements}
We would like to thank Ruslan I. Mukhamadiarov, Riya Nandi, and Priyanka for fruitful discussions. 
The authors are also indebted to Mohamed Swailem for a careful reading of the manuscript draft. 
This research was supported by the U.S. Department of Energy, Office of Basic Energy Sciences, Division of Materials Science
and Engineering under Award DE-SC0002308. 
\end{acknowledgements}

\appendix

\section{One-loop results for useful correlation functions}

In this Appendix, we briefly provide more details and intermediate results for the explicit computation of various correlation 
functions, from which one may obtain the response functions, to one-loop order in the dynamical perturbation expansion.
On the tree (mean-field) level, the non-vanishing propagators of the field theory read
\begin{equation}
\begin{aligned}
 \langle \pi_i\tilde{\pi}_j\rangle^{(0)}(p,\omega)&=\delta_{ij}\frac{i\omega-D(p^2+\tilde{h})}{[\omega-\omega_+(p)][\omega-\omega_-(p)]} \ , \\
 \langle m_i\tilde{\pi}_j\rangle^{(0)}(p,\omega)&=\epsilon_{ij}\frac{-c(p^2+\tilde{h})}{[\omega-\omega_+(p)][\omega-\omega_-(p)]} \ , \\
 \langle m_i\tilde{m}_j\rangle^{(0)}(p,\omega)&=\delta_{ij}\frac{i\omega-\lambda p^2}{[\omega-\omega_+(p)][\omega-\omega_-(p)]} \ , \\
 \langle \pi_i\tilde{m}_j\rangle^{(0)}(p,\omega)&=\epsilon_{ij}\frac{-c}{[\omega-\omega_+(p)][\omega-\omega_-(p)]}\ , \\
 \langle m_3\tilde{m}_3\rangle^{(0)}(p,\omega)&=\frac{1}{-i\omega+\Gamma p^2} \ ,
\end{aligned}
\end{equation}
where we have used the spin wave dispersions (\ref{dispersion1}).
In our convention, time and hence momentum always flow from right to left in Feynman diagrams. 

\begin{figure}[h]
    \centering
    \includegraphics[scale=0.51]{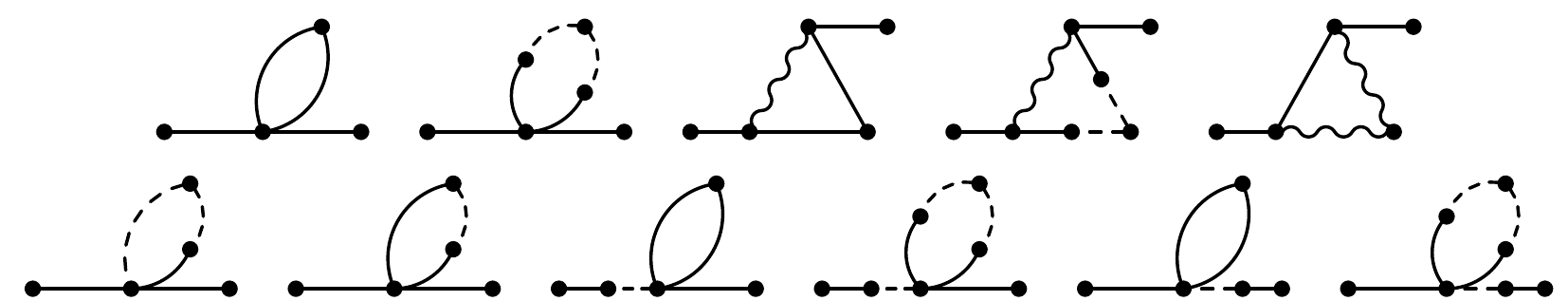}
    \caption{One-loop order contributions to $\langle\pi_i\tilde{\pi}_j\rangle$.}
    \label{pp propagator}
    \includegraphics[scale=0.7]{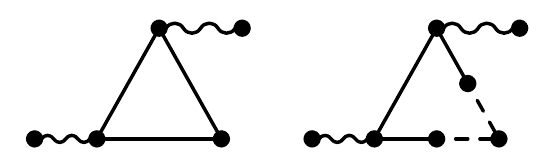}
    \caption{One-loop order contributions to $\langle m_3\tilde{m}_3\rangle$.}
    \label{m3prop}
\end{figure}
All one-loop fluctuation corrections to the propagators of the order parameter $\vec{\pi}$ and the propagator of the parallel 
component of the magnetization density $m_3$ are plotted in Figs.~\ref{pp propagator} and \ref{m3prop}. 
Here, the $\pi$, $m$, and $m_3$ fields are represented by solid, dashed, and wavy lines, respectively.
The one-loop graphs for the propagator of the perpendicular components of the magnetization density $m_1, m_2$ and the
mixed $\pi_i m_j$ propagators have the same loop structures, but just carry different external legs. 
The explicit results for the one-loop fluctuation corrections to the propagators read
\begin{widetext}
\begin{equation}
\begin{aligned}
   \langle m_i\tilde{m}_j\rangle^{(1)}(p,\omega)=&\delta_{ij}\frac{-c^2(p^2+\tilde{h})}{[\omega-\omega_+(p)]^2
   [\omega-\omega_-(p)]^2}\Bigg[g^2(p^2+\tilde{h})B(p,\omega)+(i\omega-Dp^2-D\tilde{h})uA\Bigg]\ , \\
   \langle m_i\tilde{\pi}_j\rangle^{(1)}(p,\omega)=&\epsilon_{ij}\frac{-c(p^2+\tilde{h})(i\omega-\lambda p^2)}
   {[\omega-\omega_+(p)]^2[\omega-\omega_-(p)]^2}\Bigg[g^2(p^2+\tilde{h})B(p,\omega)
   +(i\omega-Dp^2-D\tilde{h})uA\Bigg]\ , \\
   \langle\pi_i\tilde{\pi}_j\rangle^{(1)}(p,\omega)=&\delta_{ij}\frac{(i\omega-\lambda p^2)^2}
   {[\omega-\omega_+(p)]^2[\omega-\omega_-(p)]^2}\Bigg[g^2(p^2+\tilde{h})B(p,\omega)
   +(i\omega-Dp^2-D\tilde{h})uA\Bigg]\ , \\
   \langle\pi_i\tilde{m}_j\rangle^{(1)}(p,\omega)=&\epsilon_{ij}\frac{-c(i\omega-\lambda p^2)}
   {[\omega-\omega_+(p)]^2[\omega-\omega_-(p)]^2}\Bigg[g^2(p^2+\tilde{h})B(p,\omega)
  +(i\omega-Dp^2-D\tilde{h})uA\Bigg]\ , \\
   \langle m_3\tilde{m}_3\rangle^{(1)}(p,\omega)=&\frac{2g^2}{\left(-i\omega+\Gamma p^2\right)^2}\,E(p,\omega) \ ,
\end{aligned}
\end{equation}
where
\begin{equation}
\begin{aligned}
    A=&\int_k\frac{1}{k^2+\tilde{h}} \ , \\
    B(p,\omega)=&\int_k\frac{i\omega-\Gamma(p-k)^2-\lambda k^2}{(k^2+\tilde{h})
    [-i\omega+i\omega_+(k)+\Gamma(p-k)^2][-i\omega+i\omega_-(k)+\Gamma(p-k)^2]} \ , \\
    E(p,\omega)=&-\int_k\frac{[k^2-(p-k)^2][i\omega-\lambda k^2-D(p-k)^2-D\tilde{h}]}{[k^2+\tilde{h}]
    [(p-k)^2-\tilde{h}][\omega-\omega_+(k)-\omega_+(p-k)]} \\
    &\times\frac{\left([i\omega-\lambda(p-k)^2-D\tilde{h}][i\omega-\lambda k^2-\lambda(p-k)^2]+2c^2(k^2+\tilde{h})
    \right)}{[\omega-\omega_-(k)-\omega_+(p-k)][\omega-\omega_-(k)-\omega_-(p-k)]
    [\omega-\omega_+(k)-\omega_-(p-k)]} \ .
\end{aligned}
\end{equation}
\end{widetext}

\begin{figure}[h]
\centering
    \includegraphics[scale=0.6]{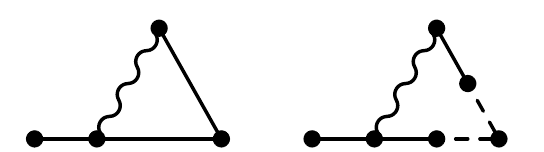}
    \caption{One-loop order contributions to $\langle\pi_i[\tilde{m}_3\pi_j]\rangle$.}
    \label{composite1}
    \includegraphics[scale=0.6]{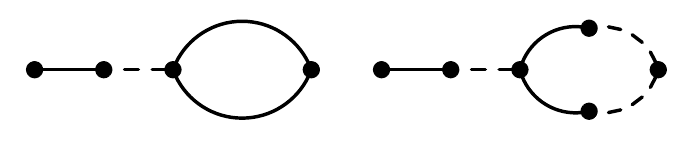}
    \caption{One-loop order contributions to $\langle\pi_i[\tilde{m}_j\vec{\pi}^2]\rangle$.}
    \label{composite2}
    \includegraphics[scale=0.6]{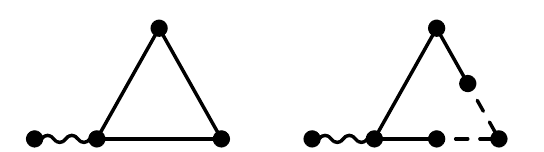}
    \caption{One-loop order contributions to $\langle m_3[\tilde{\pi}_i\pi_j]\rangle$.}
    \label{composite3}
\end{figure}

The following relations are obtained by means of explicit calculation, and lead to immediate cancellations in the evaluation of
response functions,
\begin{equation}
\begin{aligned}
    \langle m_i[\tilde{m}_jm_3]\rangle(p,\omega) &= \langle m_i[\tilde{m}_3m_j]\rangle(p,\omega) \ , \\
    \langle m_i[\tilde{\pi}_k\pi_k\pi_j]\rangle(p,\omega) &= \frac{1}{2}\langle m_i[\tilde{\pi}_j\vec{\pi}^2]
    \rangle(p,\omega) \ .
\end{aligned}
\end{equation}
The remaining one-loop results for other composite operator correlation functions that appear in the expressions for the dynamical response functions (\ref{susc}) are
\begin{equation}
\begin{aligned}
   &\langle\pi_i[\tilde{m}_3\pi_j]\rangle(p,\omega)=\epsilon_{ij}uc\frac{i\omega-\lambda p^2}{[\omega-\omega_+(p)]  
   [\omega-\omega_-(p)]} \, B(p,\omega) \ , \\
   &\langle\pi_i[\tilde{m}_j\vec{\pi}^2]\rangle(p,\omega)=\delta_{ij}\frac{2uc}{[\omega-\omega_+(p)]
   [\omega-\omega_-(p)]} \, A \ , \\
   &\langle m_3[\tilde{\pi}_i\pi_j]\rangle(p,\omega)=\epsilon_{ij}\frac{uc}{-i\omega+\Gamma p^2} \, E(p,\omega) \ ,
\end{aligned}
\end{equation}
and the associated Feynman diagrams are shown in Figs.~\ref{composite1}, \ref{composite2}, and \ref{composite3}.

At last, we may assemble these building blocks to construct the dynamical susceptibilities to one-loop order. 
The response function for the order parameter components becomes
\begin{widetext}
\begin{equation}
\begin{aligned}
    \chi_{ij}(p,\omega) &= \delta_{ij} \, \frac{D (i\omega - \lambda p^2) - c^2}{[\omega - \omega_+(p)] \,
    [\omega - \omega_-(p)]} \bigg[ 1 + \frac{(i \omega - \lambda p^2) \, (i \omega - D p^2 -D \tilde{h})}
    {[\omega - \omega_+(p)] \, [\omega - \omega_-(p)]} \, u A \biggr] \\
    &\quad \, + \delta_{ij} \, \frac{u c^2}{[\omega - \omega_+(p)] \, [\omega - \omega_-(p)]} \bigg[ A + 
    \frac{i \omega (i \omega - \lambda p^2)^2}{[\omega - \omega_+(p)] \, [\omega - \omega_-(p)]} \, B(p,\omega) \bigg] 
    \ ,
\end{aligned}
\end{equation}
while the response function for the perpendicular components of the magnetization density reads
\begin{equation}
\begin{aligned}
    R_{ij}(p,\omega) = \delta_{ij} \, \frac{-D \lambda p^2 \, (p^2 + \tilde{h}) - c^2 \, (p^2+\tilde{h}) + i \omega   
    \lambda p^2}{[\omega - \omega_+(p)] \, [\omega - \omega_-(p)]} + \delta_{ij} \, 
    \frac{-i \omega c^2 \, (p^2 + \tilde{h})}{[\omega - \omega_+(p)]^2 [\omega - \omega_-(p)]^2} \\ 
    \times \left[ (i \omega - D p^2 - D \tilde{h}) \, u A + (p^2 + \tilde{h}) u c^2 \, B(p,\omega) \right] \ ,
\end{aligned}
\end{equation}
and the response function for the parallel magnetization density component is
\begin{equation}
    R_\parallel(p,\omega) = \frac{1}{-i \omega + \Gamma p^2} \left[ \Gamma p^2 - 2 g^2 \, E(p,\omega) \right]
    - \frac{2 g^2 \, \Gamma p^2}{(-i \omega + \Gamma p^2)^2} \, E(p,\omega) \ .
\end{equation}
Straightforward algebra then yields the expressions (\ref{response}), (\ref{1-loop results}) to first order in $u$.
\end{widetext}

\section{Nonlinear sigma model: other dynamical universality classes}

In this Appendix, we briefly analyze other standard dynamical critical universality classes in the context of the nonlinear 
sigma model \cite{hohenberg1977theory,folk2006critical,tauber2014critical}. 
We shall argue that in fact all other types of universal dynamics imposed on this system turn out either trivial or incompatible
with the rigid nonlinear sigma model constraint.

\subsection{Model C dynamics}

In model C dynamics, one allows for the static coupling of the conserved energy density $\mathcal{E}(x,t)$ to the order
parameter. 
With the thermodynamics of a critical system captured by the Landau--Ginzburg--Wilson effective Hamiltonian, the energy
density becomes directly coupled to the square of order parameter, whence the two-point correlation function for the energy
density fluctuations $\langle \mathcal{E} \mathcal{E} \rangle$ is proportional to the specific heat $C_V$, as it should
\cite{hohenberg1977theory}.
Yet in the nonlinear sigma model, one imposes the rigid constraint $\vec{n}^2 = 1$. 
Coupling the energy density to $\vec{n}^2$ in the Hamiltonian through a term 
$\frac{1}{2} \mathcal{E}(x,t)^2 +  g \mathcal{E}(x,t) \, \vec{n}^2$ thus becomes trivial; the field $\mathcal{E}(x,t)$ 
may be readily integrated out without any effects on the order parameter fluctuations.

In the hope of incorporating nontrivial fluctuations of the energy density, we could instead consider the static coupling 
\begin{equation}
    H_{\mathrm{C}}=\frac{1}{2} \int d^dx \left[ \mathcal{E} - g \left( \nabla \vec{n} \right)^2 \right]^2 .
\end{equation}
Note that this term is quadratic in $\mathcal{E}$ and can be integrated out to show that $\mathcal{E}$ is proportional to 
$\frac{1}{2} \left( \nabla \vec{n} \right)^2$, i.e., the Hamiltonian density. 
However, a naive dimensional analysis then immediately yields that the resulting effective coupling $g^2 u$ has the negative
scaling dimension $-d$ and is thus irrelevant in the RG sense. 
In consequence, near the critical point, the system decouples from the energy density.
This suggests that the dynamic exponent of the order parameter assumes its model A value $z_n = 2 + c \, \eta$, while the
dynamic scaling exponent for the conserved energy density retains its Gaussian value $z_\mathcal{E} = 2$. 
This result is consistent with well-established findings from model C critical dynamics based on the $O(n)$-symmetric 
Landau--Ginzburg--Wilson theory, namely that for negative specific heat exponent $\alpha < 0$ or vector order parameter 
component number $n \geq 2$, the energy density asymptotically decouples from the order parameter in any dimension 
$d > 2$, leading to model A critical scaling for the order parameter
\cite{hohenberg1977theory,akkineni2004nonequilibrium,folk2006critical,tauber2014critical}.

\subsection{Critical dynamics with conserved order parameter}

In the standard dynamical critical universality classes with conserved total order parameter, i.e., models B, D and J, one 
requires the total order parameter in the system to be conserved \cite{hohenberg1977theory}.
In the absence of any nonlinear couplings (i.e., the Gaussian approximation), the order parameter field consequently relaxes 
diffusively.
However, to implement the constraint of the nonlinear sigma model, one strictly fixes the $\sigma$ component and
considers only fluctuations of the transverse Goldstone modes $\vec{\pi}$. 
Yet to be consistent with the conservation law for the total order parameter, one would need to properly eliminate the 
zero-wavevector fluctuations for all modes, and at finite wavevectors allow diffusive couplings between the longitudinal
$\sigma$ and the transverse $\vec{\pi}$ sectors. 
Thus, the dynamical universality classes with conserved total order parameter appear inaccessible in the context of the 
nonlinear sigma model.

\section{Scaling behavior at $d = 3.9$}

\begin{figure}[h]
\centering
\subfloat[]{\includegraphics[scale=0.35]{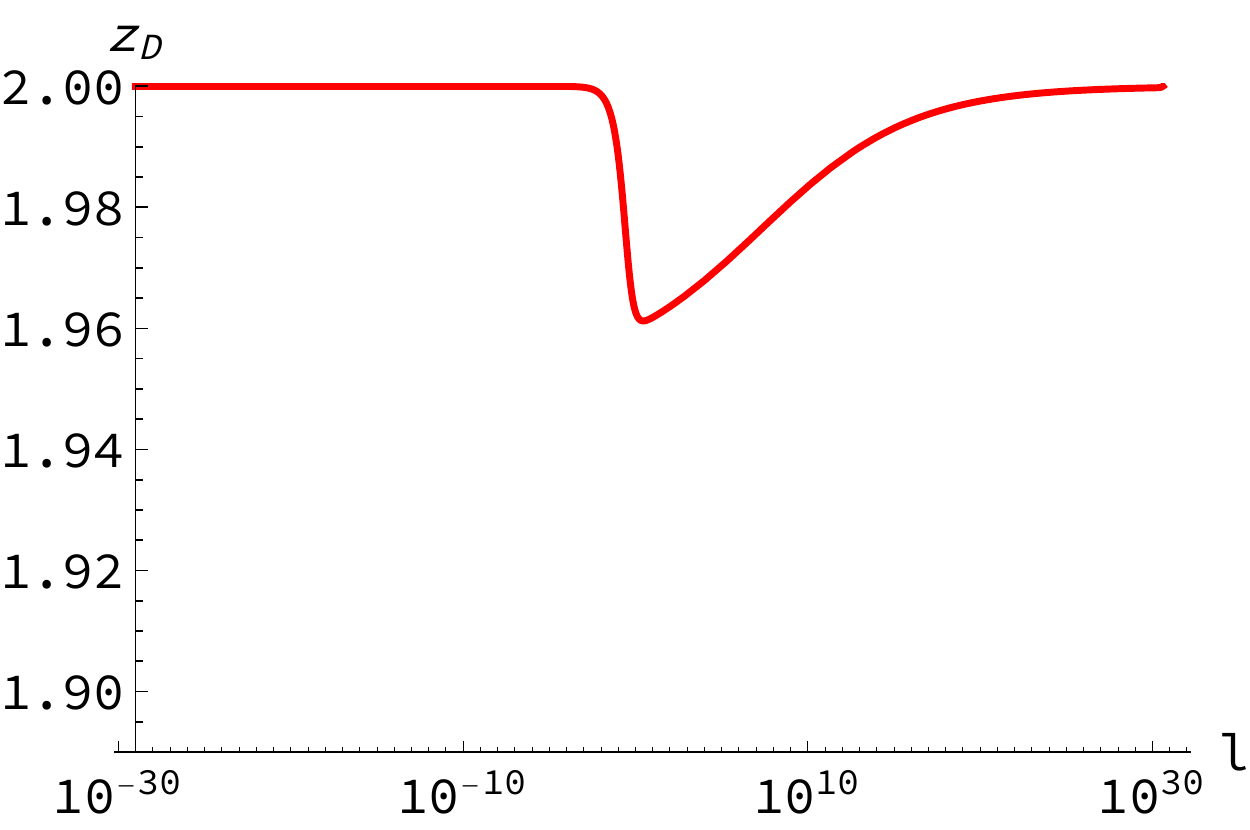}}
\subfloat[]{\includegraphics[scale=0.35]{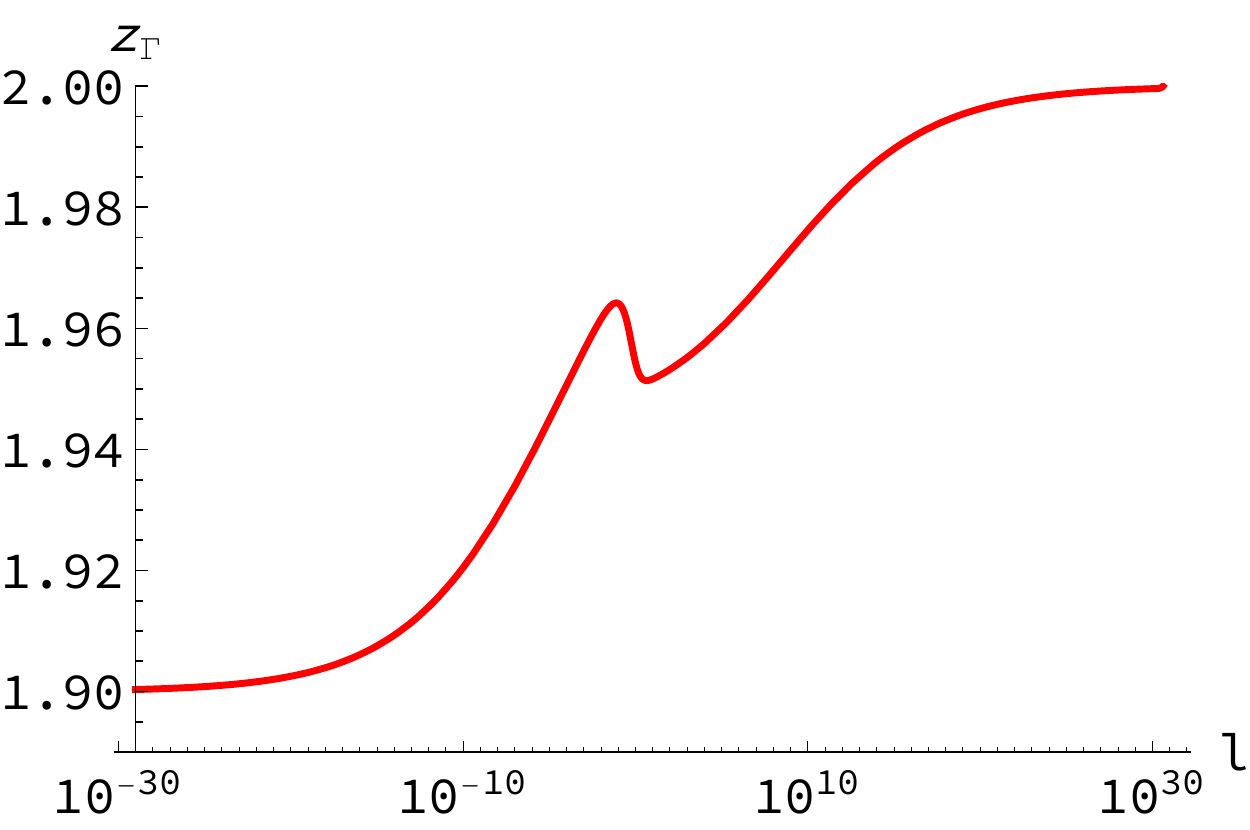}}
\caption{RG flow of the effective dynamic exponents $z_D(l)$ and $z_\Gamma(l)$ with the initial conditions $f(1) = 0.1$,
	$D(1) = 1.5$, $\Gamma(1) = 1$, $\lambda(1) = \lambda_R = 1$, and $c(1) = 1$, for $\epsilon = 0.1$ ($d = 3.9$).
      Asymptotically as $l \to 0$, $z_D = 2$ and $z_\Gamma = 2 - \epsilon$; the crossover for the latter to reach the 
	coexistence scaling regime however requires extremely small flow parameter values $l < 10^{-20}$. }
\label{z3.9}
\end{figure}

We numerically solved the coupled RG flow equations (\ref{floweqs}) for $\epsilon=0.1$ and plot the ensuing flow of the
effective dynamical exponents $z_D(l)$ and $z_\Gamma(l)$ in Fig.~\ref{z3.9}. 
While of course $d = 3.9$ is not a physically realizable spatial dimension, the $\epsilon$ expansion is on safer grounds in 
this case.
However, we observe the same qualitiative flow behavior as in three dimensions:
Starting from the Gaussian fixed point values on the right, the effective exponents display a brief crossover to their critical 
values, and subsequently approach their asymptotic ordered-phase fixed point values. 
The order parameter $\vec{\pi}$ behaves purely diffusively, $z_D = 2$, but the longitudinal magnetization component 
$m_3$ acquires nontrivial corrections due to its coupling to the transverse Goldstone modes, and ultimately displays 
sub-diffusive scaling in the coexistence limit governed by $z_\Gamma = 2 - \epsilon$.
Note that the crossover to this asymptotic coexistence scaling behavior happens exceedingly slowly for small values of 
$\epsilon$, as becomes apparent by the huge range of flow parameters $l$ required to capture the full crossover in 
Fig.~\ref{z3.9}(b).
At $d_c = 4$, this slow power law approach would be replaced by logarithmic corrections to the mean-field diffusive
relaxation for the longitudinal magnetization density.

\bibliographystyle{apsrev4-2}
\nocite{*}

\bibliography{ref}

\begin{thebibliography}{55}%
\makeatletter
\providecommand \@ifxundefined [1]{%
 \@ifx{#1\undefined}
}%
\providecommand \@ifnum [1]{%
 \ifnum #1\expandafter \@firstoftwo
 \else \expandafter \@secondoftwo
 \fi
}%
\providecommand \@ifx [1]{%
 \ifx #1\expandafter \@firstoftwo
 \else \expandafter \@secondoftwo
 \fi
}%
\providecommand \natexlab [1]{#1}%
\providecommand \enquote  [1]{``#1''}%
\providecommand \bibnamefont  [1]{#1}%
\providecommand \bibfnamefont [1]{#1}%
\providecommand \citenamefont [1]{#1}%
\providecommand \href@noop [0]{\@secondoftwo}%
\providecommand \href [0]{\begingroup \@sanitize@url \@href}%
\providecommand \@href[1]{\@@startlink{#1}\@@href}%
\providecommand \@@href[1]{\endgroup#1\@@endlink}%
\providecommand \@sanitize@url [0]{\catcode `\\12\catcode `\$12\catcode
  `\&12\catcode `\#12\catcode `\^12\catcode `\_12\catcode `\%12\relax}%
\providecommand \@@startlink[1]{}%
\providecommand \@@endlink[0]{}%
\providecommand \url  [0]{\begingroup\@sanitize@url \@url }%
\providecommand \@url [1]{\endgroup\@href {#1}{\urlprefix }}%
\providecommand \urlprefix  [0]{URL }%
\providecommand \Eprint [0]{\href }%
\providecommand \doibase [0]{https://doi.org/}%
\providecommand \selectlanguage [0]{\@gobble}%
\providecommand \bibinfo  [0]{\@secondoftwo}%
\providecommand \bibfield  [0]{\@secondoftwo}%
\providecommand \translation [1]{[#1]}%
\providecommand \BibitemOpen [0]{}%
\providecommand \bibitemStop [0]{}%
\providecommand \bibitemNoStop [0]{.\EOS\space}%
\providecommand \EOS [0]{\spacefactor3000\relax}%
\providecommand \BibitemShut  [1]{\csname bibitem#1\endcsname}%
\let\auto@bib@innerbib\@empty
\bibitem [{\citenamefont {Wilson}\ and\ \citenamefont
  {Fisher}(1972)}]{wilson1972critical}%
  \BibitemOpen
  \bibfield  {author} {\bibinfo {author} {\bibfnamefont {K.~G.}\ \bibnamefont
  {Wilson}}\ and\ \bibinfo {author} {\bibfnamefont {M.~E.}\ \bibnamefont
  {Fisher}},\ }\href@noop {} {\bibfield  {journal} {\bibinfo  {journal}
  {Physical Review Letters}\ }\textbf {\bibinfo {volume} {28}},\ \bibinfo
  {pages} {240} (\bibinfo {year} {1972})}\BibitemShut {NoStop}%
\bibitem [{\citenamefont {Wilson}\ and\ \citenamefont
  {Kogut}(1974)}]{wilson1974renormalization}%
  \BibitemOpen
  \bibfield  {author} {\bibinfo {author} {\bibfnamefont {K.~G.}\ \bibnamefont
  {Wilson}}\ and\ \bibinfo {author} {\bibfnamefont {J.}~\bibnamefont {Kogut}},\
  }\href@noop {} {\bibfield  {journal} {\bibinfo  {journal} {Physics Reports}\
  }\textbf {\bibinfo {volume} {12}},\ \bibinfo {pages} {75} (\bibinfo {year}
  {1974})}\BibitemShut {NoStop}%
\bibitem [{\citenamefont {Wilson}(1983)}]{wilson1983renormalization}%
  \BibitemOpen
  \bibfield  {author} {\bibinfo {author} {\bibfnamefont {K.~G.}\ \bibnamefont
  {Wilson}},\ }\href@noop {} {\bibfield  {journal} {\bibinfo  {journal}
  {Reviews of Modern Physics}\ }\textbf {\bibinfo {volume} {55}},\ \bibinfo
  {pages} {583} (\bibinfo {year} {1983})}\BibitemShut {NoStop}%
\bibitem [{\citenamefont {Hasselmann}\ \emph {et~al.}(2004)\citenamefont
  {Hasselmann}, \citenamefont {Ledowski},\ and\ \citenamefont
  {Kopietz}}]{hasselmann2004critical}%
  \BibitemOpen
  \bibfield  {author} {\bibinfo {author} {\bibfnamefont {N.}~\bibnamefont
  {Hasselmann}}, \bibinfo {author} {\bibfnamefont {S.}~\bibnamefont
  {Ledowski}},\ and\ \bibinfo {author} {\bibfnamefont {P.}~\bibnamefont
  {Kopietz}},\ }\href@noop {} {\bibfield  {journal} {\bibinfo  {journal}
  {Physical Review A}\ }\textbf {\bibinfo {volume} {70}},\ \bibinfo {pages}
  {063621} (\bibinfo {year} {2004})}\BibitemShut {NoStop}%
\bibitem [{\citenamefont {Stoki{\'c}}\ \emph {et~al.}(2010)\citenamefont
  {Stoki{\'c}}, \citenamefont {Friman},\ and\ \citenamefont
  {Redlich}}]{stokic2010functional}%
  \BibitemOpen
  \bibfield  {author} {\bibinfo {author} {\bibfnamefont {B.}~\bibnamefont
  {Stoki{\'c}}}, \bibinfo {author} {\bibfnamefont {B.}~\bibnamefont {Friman}},\
  and\ \bibinfo {author} {\bibfnamefont {K.}~\bibnamefont {Redlich}},\
  }\href@noop {} {\bibfield  {journal} {\bibinfo  {journal} {The European
  Physical Journal C}\ }\textbf {\bibinfo {volume} {67}},\ \bibinfo {pages}
  {425} (\bibinfo {year} {2010})}\BibitemShut {NoStop}%
\bibitem [{\citenamefont {Litim}\ and\ \citenamefont
  {Zappala}(2011)}]{litim2011ising}%
  \BibitemOpen
  \bibfield  {author} {\bibinfo {author} {\bibfnamefont {D.~F.}\ \bibnamefont
  {Litim}}\ and\ \bibinfo {author} {\bibfnamefont {D.}~\bibnamefont
  {Zappala}},\ }\href@noop {} {\bibfield  {journal} {\bibinfo  {journal}
  {Physical Review D}\ }\textbf {\bibinfo {volume} {83}},\ \bibinfo {pages}
  {085009} (\bibinfo {year} {2011})}\BibitemShut {NoStop}%
\bibitem [{\citenamefont {El-Showk}\ \emph {et~al.}(2012)\citenamefont
  {El-Showk}, \citenamefont {Paulos}, \citenamefont {Poland}, \citenamefont
  {Rychkov}, \citenamefont {Simmons-Duffin},\ and\ \citenamefont
  {Vichi}}]{el2012solving}%
  \BibitemOpen
  \bibfield  {author} {\bibinfo {author} {\bibfnamefont {S.}~\bibnamefont
  {El-Showk}}, \bibinfo {author} {\bibfnamefont {M.~F.}\ \bibnamefont
  {Paulos}}, \bibinfo {author} {\bibfnamefont {D.}~\bibnamefont {Poland}},
  \bibinfo {author} {\bibfnamefont {S.}~\bibnamefont {Rychkov}}, \bibinfo
  {author} {\bibfnamefont {D.}~\bibnamefont {Simmons-Duffin}},\ and\ \bibinfo
  {author} {\bibfnamefont {A.}~\bibnamefont {Vichi}},\ }\href@noop {}
  {\bibfield  {journal} {\bibinfo  {journal} {Physical Review D}\ }\textbf
  {\bibinfo {volume} {86}},\ \bibinfo {pages} {025022} (\bibinfo {year}
  {2012})}\BibitemShut {NoStop}%
\bibitem [{\citenamefont {Poland}\ and\ \citenamefont
  {Simmons-Duffin}(2016)}]{poland2016conformal}%
  \BibitemOpen
  \bibfield  {author} {\bibinfo {author} {\bibfnamefont {D.}~\bibnamefont
  {Poland}}\ and\ \bibinfo {author} {\bibfnamefont {D.}~\bibnamefont
  {Simmons-Duffin}},\ }\href@noop {} {\bibfield  {journal} {\bibinfo  {journal}
  {Nature Physics}\ }\textbf {\bibinfo {volume} {12}},\ \bibinfo {pages} {535}
  (\bibinfo {year} {2016})}\BibitemShut {NoStop}%
\bibitem [{\citenamefont {Poland}\ \emph {et~al.}(2019)\citenamefont {Poland},
  \citenamefont {Rychkov},\ and\ \citenamefont {Vichi}}]{poland2019conformal}%
  \BibitemOpen
  \bibfield  {author} {\bibinfo {author} {\bibfnamefont {D.}~\bibnamefont
  {Poland}}, \bibinfo {author} {\bibfnamefont {S.}~\bibnamefont {Rychkov}},\
  and\ \bibinfo {author} {\bibfnamefont {A.}~\bibnamefont {Vichi}},\
  }\href@noop {} {\bibfield  {journal} {\bibinfo  {journal} {Reviews of Modern
  Physics}\ }\textbf {\bibinfo {volume} {91}},\ \bibinfo {pages} {015002}
  (\bibinfo {year} {2019})}\BibitemShut {NoStop}%
\bibitem [{\citenamefont {Hohenberg}\ and\ \citenamefont
  {Halperin}(1977)}]{hohenberg1977theory}%
  \BibitemOpen
  \bibfield  {author} {\bibinfo {author} {\bibfnamefont {P.~C.}\ \bibnamefont
  {Hohenberg}}\ and\ \bibinfo {author} {\bibfnamefont {B.~I.}\ \bibnamefont
  {Halperin}},\ }\href@noop {} {\bibfield  {journal} {\bibinfo  {journal}
  {Reviews of Modern Physics}\ }\textbf {\bibinfo {volume} {49}},\ \bibinfo
  {pages} {435} (\bibinfo {year} {1977})}\BibitemShut {NoStop}%
\bibitem [{\citenamefont {Ferrell}\ \emph {et~al.}(1967)\citenamefont
  {Ferrell}, \citenamefont {Menyhard}, \citenamefont {Schmidt}, \citenamefont
  {Schwabl},\ and\ \citenamefont {Sz{\'e}pfalusy}}]{ferrell1967dispersion}%
  \BibitemOpen
  \bibfield  {author} {\bibinfo {author} {\bibfnamefont {R.~A.}\ \bibnamefont
  {Ferrell}}, \bibinfo {author} {\bibfnamefont {N.}~\bibnamefont {Menyhard}},
  \bibinfo {author} {\bibfnamefont {H.}~\bibnamefont {Schmidt}}, \bibinfo
  {author} {\bibfnamefont {F.}~\bibnamefont {Schwabl}},\ and\ \bibinfo {author}
  {\bibfnamefont {P.}~\bibnamefont {Sz{\'e}pfalusy}},\ }\href@noop {}
  {\bibfield  {journal} {\bibinfo  {journal} {Physical Review Letters}\
  }\textbf {\bibinfo {volume} {18}},\ \bibinfo {pages} {891} (\bibinfo {year}
  {1967})}\BibitemShut {NoStop}%
\bibitem [{\citenamefont {Halperin}\ and\ \citenamefont
  {Hohenberg}(1969)}]{halperin1969scaling}%
  \BibitemOpen
  \bibfield  {author} {\bibinfo {author} {\bibfnamefont {B.~I.}\ \bibnamefont
  {Halperin}}\ and\ \bibinfo {author} {\bibfnamefont {P.~C.}\ \bibnamefont
  {Hohenberg}},\ }\href@noop {} {\bibfield  {journal} {\bibinfo  {journal}
  {Physical Review}\ }\textbf {\bibinfo {volume} {177}},\ \bibinfo {pages}
  {952} (\bibinfo {year} {1969})}\BibitemShut {NoStop}%
\bibitem [{\citenamefont {Fixman}(1962)}]{fixman1962viscosity}%
  \BibitemOpen
  \bibfield  {author} {\bibinfo {author} {\bibfnamefont {M.}~\bibnamefont
  {Fixman}},\ }\href@noop {} {\bibfield  {journal} {\bibinfo  {journal} {The
  Journal of Chemical Physics}\ }\textbf {\bibinfo {volume} {36}},\ \bibinfo
  {pages} {310} (\bibinfo {year} {1962})}\BibitemShut {NoStop}%
\bibitem [{\citenamefont {Kadanoff}\ and\ \citenamefont
  {Swift}(1968)}]{kadanoff1968transport}%
  \BibitemOpen
  \bibfield  {author} {\bibinfo {author} {\bibfnamefont {L.~P.}\ \bibnamefont
  {Kadanoff}}\ and\ \bibinfo {author} {\bibfnamefont {J.}~\bibnamefont
  {Swift}},\ }\href@noop {} {\bibfield  {journal} {\bibinfo  {journal}
  {Physical Review}\ }\textbf {\bibinfo {volume} {166}},\ \bibinfo {pages} {89}
  (\bibinfo {year} {1968})}\BibitemShut {NoStop}%
\bibitem [{\citenamefont {Kawasaki}(1967)}]{kawasaki1967anomalous}%
  \BibitemOpen
  \bibfield  {author} {\bibinfo {author} {\bibfnamefont {K.}~\bibnamefont
  {Kawasaki}},\ }\href@noop {} {\bibfield  {journal} {\bibinfo  {journal}
  {Journal of Physics and Chemistry of Solids}\ }\textbf {\bibinfo {volume}
  {28}},\ \bibinfo {pages} {1277} (\bibinfo {year} {1967})}\BibitemShut
  {NoStop}%
\bibitem [{\citenamefont {Kawasaki}(1970)}]{kawasaki1970kinetic}%
  \BibitemOpen
  \bibfield  {author} {\bibinfo {author} {\bibfnamefont {K.}~\bibnamefont
  {Kawasaki}},\ }\href@noop {} {\bibfield  {journal} {\bibinfo  {journal}
  {Annals of Physics}\ }\textbf {\bibinfo {volume} {61}},\ \bibinfo {pages} {1}
  (\bibinfo {year} {1970})}\BibitemShut {NoStop}%
\bibitem [{\citenamefont {Gunton}\ and\ \citenamefont
  {Kawasaki}(1976)}]{gunton1976renormalization}%
  \BibitemOpen
  \bibfield  {author} {\bibinfo {author} {\bibfnamefont {J.~D.}\ \bibnamefont
  {Gunton}}\ and\ \bibinfo {author} {\bibfnamefont {K.}~\bibnamefont
  {Kawasaki}},\ }\href@noop {} {\bibfield  {journal} {\bibinfo  {journal}
  {Progress of Theoretical Physics}\ }\textbf {\bibinfo {volume} {56}},\
  \bibinfo {pages} {61} (\bibinfo {year} {1976})}\BibitemShut {NoStop}%
\bibitem [{\citenamefont {Vasil'ev}(2004)}]{vasil2004field}%
  \BibitemOpen
  \bibfield  {author} {\bibinfo {author} {\bibfnamefont {A.~N.}\ \bibnamefont
  {Vasil'ev}},\ }\href@noop {} {\emph {\bibinfo {title} {The field theoretic
  renormalization group in critical behavior theory and stochastic dynamics}}}\
  (\bibinfo  {publisher} {Chapman and Hall/CRC},\ \bibinfo {year}
  {2004})\BibitemShut {NoStop}%
\bibitem [{\citenamefont {Folk}\ and\ \citenamefont
  {Moser}(2006)}]{folk2006critical}%
  \BibitemOpen
  \bibfield  {author} {\bibinfo {author} {\bibfnamefont {R.}~\bibnamefont
  {Folk}}\ and\ \bibinfo {author} {\bibfnamefont {G.}~\bibnamefont {Moser}},\
  }\href@noop {} {\bibfield  {journal} {\bibinfo  {journal} {Journal of Physics
  A: Mathematical and General}\ }\textbf {\bibinfo {volume} {39}},\ \bibinfo
  {pages} {R207} (\bibinfo {year} {2006})}\BibitemShut {NoStop}%
\bibitem [{\citenamefont {T{\"a}uber}(2014)}]{tauber2014critical}%
  \BibitemOpen
  \bibfield  {author} {\bibinfo {author} {\bibfnamefont {U.~C.}\ \bibnamefont
  {T{\"a}uber}},\ }\href@noop {} {\emph {\bibinfo {title} {Critical dynamics: a
  field theory approach to equilibrium and non-equilibrium scaling behavior}}}\
  (\bibinfo  {publisher} {Cambridge University Press},\ \bibinfo {year}
  {2014})\BibitemShut {NoStop}%
\bibitem [{\citenamefont {Halperin}\ \emph {et~al.}(1974)\citenamefont
  {Halperin}, \citenamefont {Hohenberg},\ and\ \citenamefont
  {Siggia}}]{halperin1974renormalization}%
  \BibitemOpen
  \bibfield  {author} {\bibinfo {author} {\bibfnamefont {B.}~\bibnamefont
  {Halperin}}, \bibinfo {author} {\bibfnamefont {P.~C.}\ \bibnamefont
  {Hohenberg}},\ and\ \bibinfo {author} {\bibfnamefont {E.~D.}\ \bibnamefont
  {Siggia}},\ }\href@noop {} {\bibfield  {journal} {\bibinfo  {journal}
  {Physical Review Letters}\ }\textbf {\bibinfo {volume} {32}},\ \bibinfo
  {pages} {1289} (\bibinfo {year} {1974})}\BibitemShut {NoStop}%
\bibitem [{\citenamefont {Ma}\ and\ \citenamefont
  {Mazenko}(1974)}]{ma1974critical}%
  \BibitemOpen
  \bibfield  {author} {\bibinfo {author} {\bibfnamefont {S.~K.}\ \bibnamefont
  {Ma}}\ and\ \bibinfo {author} {\bibfnamefont {G.~F.}\ \bibnamefont
  {Mazenko}},\ }\href@noop {} {\bibfield  {journal} {\bibinfo  {journal}
  {Physical Review Letters}\ }\textbf {\bibinfo {volume} {33}},\ \bibinfo
  {pages} {1383} (\bibinfo {year} {1974})}\BibitemShut {NoStop}%
\bibitem [{\citenamefont {Sasv{\'a}ri}\ \emph {et~al.}(1975)\citenamefont
  {Sasv{\'a}ri}, \citenamefont {Schwabl},\ and\ \citenamefont
  {Sz{\'e}pfalusy}}]{sasvari1975hydrodynamics}%
  \BibitemOpen
  \bibfield  {author} {\bibinfo {author} {\bibfnamefont {L.}~\bibnamefont
  {Sasv{\'a}ri}}, \bibinfo {author} {\bibfnamefont {F.}~\bibnamefont
  {Schwabl}},\ and\ \bibinfo {author} {\bibfnamefont {P.}~\bibnamefont
  {Sz{\'e}pfalusy}},\ }\href@noop {} {\bibfield  {journal} {\bibinfo  {journal}
  {Physica A: Statistical Mechanics and its Applications}\ }\textbf {\bibinfo
  {volume} {81}},\ \bibinfo {pages} {108} (\bibinfo {year} {1975})}\BibitemShut
  {NoStop}%
\bibitem [{\citenamefont {Halperin}\ \emph {et~al.}(1980)\citenamefont
  {Halperin}, \citenamefont {Hohenberg},\ and\ \citenamefont
  {Siggia}}]{halperin1980erratum}%
  \BibitemOpen
  \bibfield  {author} {\bibinfo {author} {\bibfnamefont {B.~I.}\ \bibnamefont
  {Halperin}}, \bibinfo {author} {\bibfnamefont {P.~C.}\ \bibnamefont
  {Hohenberg}},\ and\ \bibinfo {author} {\bibfnamefont {E.~D.}\ \bibnamefont
  {Siggia}},\ }\href@noop {} {\bibfield  {journal} {\bibinfo  {journal}
  {Physical Review B}\ }\textbf {\bibinfo {volume} {21}},\ \bibinfo {pages}
  {2044} (\bibinfo {year} {1980})}\BibitemShut {NoStop}%
\bibitem [{\citenamefont {Sasv{\'a}ri}\ and\ \citenamefont
  {Sz{\'e}pfalusy}(1977)}]{sasvari1977dynamic}%
  \BibitemOpen
  \bibfield  {author} {\bibinfo {author} {\bibfnamefont {L.}~\bibnamefont
  {Sasv{\'a}ri}}\ and\ \bibinfo {author} {\bibfnamefont {P.}~\bibnamefont
  {Sz{\'e}pfalusy}},\ }\href@noop {} {\bibfield  {journal} {\bibinfo  {journal}
  {Physica A: Statistical Mechanics and its Applications}\ }\textbf {\bibinfo
  {volume} {87}},\ \bibinfo {pages} {1} (\bibinfo {year} {1977})}\BibitemShut
  {NoStop}%
\bibitem [{\citenamefont {De~Dominicis}\ and\ \citenamefont
  {Peliti}(1978)}]{de1978field}%
  \BibitemOpen
  \bibfield  {author} {\bibinfo {author} {\bibfnamefont {C.}~\bibnamefont
  {De~Dominicis}}\ and\ \bibinfo {author} {\bibfnamefont {L.}~\bibnamefont
  {Peliti}},\ }\href@noop {} {\bibfield  {journal} {\bibinfo  {journal}
  {Physical Review B}\ }\textbf {\bibinfo {volume} {18}},\ \bibinfo {pages}
  {353} (\bibinfo {year} {1978})}\BibitemShut {NoStop}%
\bibitem [{\citenamefont {Coldea}\ \emph {et~al.}(1998)\citenamefont {Coldea},
  \citenamefont {Cowley}, \citenamefont {Perring}, \citenamefont {McMorrow},\
  and\ \citenamefont {Roessli}}]{coldea1998critical}%
  \BibitemOpen
  \bibfield  {author} {\bibinfo {author} {\bibfnamefont {R.}~\bibnamefont
  {Coldea}}, \bibinfo {author} {\bibfnamefont {R.~A.}\ \bibnamefont {Cowley}},
  \bibinfo {author} {\bibfnamefont {T.~G.}\ \bibnamefont {Perring}}, \bibinfo
  {author} {\bibfnamefont {D.~F.}\ \bibnamefont {McMorrow}},\ and\ \bibinfo
  {author} {\bibfnamefont {B.}~\bibnamefont {Roessli}},\ }\href@noop {}
  {\bibfield  {journal} {\bibinfo  {journal} {Physical Review B}\ }\textbf
  {\bibinfo {volume} {57}},\ \bibinfo {pages} {5281} (\bibinfo {year}
  {1998})}\BibitemShut {NoStop}%
\bibitem [{\citenamefont {Bunker}\ \emph {et~al.}(1996)\citenamefont {Bunker},
  \citenamefont {Chen},\ and\ \citenamefont {Landau}}]{bunker1996critical}%
  \BibitemOpen
  \bibfield  {author} {\bibinfo {author} {\bibfnamefont {A.}~\bibnamefont
  {Bunker}}, \bibinfo {author} {\bibfnamefont {K.}~\bibnamefont {Chen}},\ and\
  \bibinfo {author} {\bibfnamefont {D.~P.}\ \bibnamefont {Landau}},\
  }\href@noop {} {\bibfield  {journal} {\bibinfo  {journal} {Physical Review
  B}\ }\textbf {\bibinfo {volume} {54}},\ \bibinfo {pages} {9259} (\bibinfo
  {year} {1996})}\BibitemShut {NoStop}%
\bibitem [{\citenamefont {Tsai}\ and\ \citenamefont
  {Landau}(2003)}]{tsai2003critical}%
  \BibitemOpen
  \bibfield  {author} {\bibinfo {author} {\bibfnamefont {S.-H.}\ \bibnamefont
  {Tsai}}\ and\ \bibinfo {author} {\bibfnamefont {D.~P.}\ \bibnamefont
  {Landau}},\ }\href@noop {} {\bibfield  {journal} {\bibinfo  {journal}
  {Physical Review B}\ }\textbf {\bibinfo {volume} {67}},\ \bibinfo {pages}
  {104411} (\bibinfo {year} {2003})}\BibitemShut {NoStop}%
\bibitem [{\citenamefont {Nandi}\ and\ \citenamefont
  {T{\"a}uber}(2020)}]{nandi2020critical}%
  \BibitemOpen
  \bibfield  {author} {\bibinfo {author} {\bibfnamefont {R.}~\bibnamefont
  {Nandi}}\ and\ \bibinfo {author} {\bibfnamefont {U.~C.}\ \bibnamefont
  {T{\"a}uber}},\ }\href@noop {} {\bibfield  {journal} {\bibinfo  {journal}
  {Physical Review E}\ }\textbf {\bibinfo {volume} {102}},\ \bibinfo {pages}
  {052114} (\bibinfo {year} {2020})}\BibitemShut {NoStop}%
\bibitem [{\citenamefont {Brezin}\ and\ \citenamefont
  {Zinn-Justin}(1976)}]{brezin1976renormalization}%
  \BibitemOpen
  \bibfield  {author} {\bibinfo {author} {\bibfnamefont {E.}~\bibnamefont
  {Brezin}}\ and\ \bibinfo {author} {\bibfnamefont {J.}~\bibnamefont
  {Zinn-Justin}},\ }\href@noop {} {\bibfield  {journal} {\bibinfo  {journal}
  {Physical Review Letters}\ }\textbf {\bibinfo {volume} {36}},\ \bibinfo
  {pages} {691} (\bibinfo {year} {1976})}\BibitemShut {NoStop}%
\bibitem [{\citenamefont {Chakravarty}\ \emph {et~al.}(1989)\citenamefont
  {Chakravarty}, \citenamefont {Halperin},\ and\ \citenamefont
  {Nelson}}]{chakravarty1989two}%
  \BibitemOpen
  \bibfield  {author} {\bibinfo {author} {\bibfnamefont {S.}~\bibnamefont
  {Chakravarty}}, \bibinfo {author} {\bibfnamefont {B.~I.}\ \bibnamefont
  {Halperin}},\ and\ \bibinfo {author} {\bibfnamefont {D.~R.}\ \bibnamefont
  {Nelson}},\ }\href@noop {} {\bibfield  {journal} {\bibinfo  {journal}
  {Physical Review B}\ }\textbf {\bibinfo {volume} {39}},\ \bibinfo {pages}
  {2344} (\bibinfo {year} {1989})}\BibitemShut {NoStop}%
\bibitem [{\citenamefont {Bausch}\ \emph {et~al.}(1980)\citenamefont {Bausch},
  \citenamefont {Janssen},\ and\ \citenamefont
  {Yamazaki}}]{bausch1980nonlinear}%
  \BibitemOpen
  \bibfield  {author} {\bibinfo {author} {\bibfnamefont {R.}~\bibnamefont
  {Bausch}}, \bibinfo {author} {\bibfnamefont {H.~K.}\ \bibnamefont
  {Janssen}},\ and\ \bibinfo {author} {\bibfnamefont {Y.}~\bibnamefont
  {Yamazaki}},\ }\href@noop {} {\bibfield  {journal} {\bibinfo  {journal}
  {Zeitschrift f{\"u}r Physik B Condensed Matter}\ }\textbf {\bibinfo {volume}
  {37}},\ \bibinfo {pages} {163} (\bibinfo {year} {1980})}\BibitemShut
  {NoStop}%
\bibitem [{\citenamefont {Fedorenko}\ and\ \citenamefont
  {Trimper}(2006)}]{fedorenko2006critical}%
  \BibitemOpen
  \bibfield  {author} {\bibinfo {author} {\bibfnamefont {A.~A.}\ \bibnamefont
  {Fedorenko}}\ and\ \bibinfo {author} {\bibfnamefont {S.}~\bibnamefont
  {Trimper}},\ }\href@noop {} {\bibfield  {journal} {\bibinfo  {journal} {EPL
  (Europhysics Letters)}\ }\textbf {\bibinfo {volume} {74}},\ \bibinfo {pages}
  {89} (\bibinfo {year} {2006})}\BibitemShut {NoStop}%
\bibitem [{\citenamefont {Janssen}(1976)}]{janssen1976lagrangean}%
  \BibitemOpen
  \bibfield  {author} {\bibinfo {author} {\bibfnamefont {H.~K.}\ \bibnamefont
  {Janssen}},\ }\href@noop {} {\bibfield  {journal} {\bibinfo  {journal}
  {Zeitschrift f{\"u}r Physik B Condensed Matter}\ }\textbf {\bibinfo {volume}
  {23}},\ \bibinfo {pages} {377} (\bibinfo {year} {1976})}\BibitemShut
  {NoStop}%
\bibitem [{\citenamefont {De~Dominicis}\ \emph {et~al.}(1975)\citenamefont
  {De~Dominicis}, \citenamefont {Br{\'e}zin},\ and\ \citenamefont
  {Zinn-Justin}}]{de1975field}%
  \BibitemOpen
  \bibfield  {author} {\bibinfo {author} {\bibfnamefont {C.}~\bibnamefont
  {De~Dominicis}}, \bibinfo {author} {\bibfnamefont {E.}~\bibnamefont
  {Br{\'e}zin}},\ and\ \bibinfo {author} {\bibfnamefont {J.}~\bibnamefont
  {Zinn-Justin}},\ }\href@noop {} {\bibfield  {journal} {\bibinfo  {journal}
  {Physical Review B}\ }\textbf {\bibinfo {volume} {12}},\ \bibinfo {pages}
  {4945} (\bibinfo {year} {1975})}\BibitemShut {NoStop}%
\bibitem [{\citenamefont {Br{\'e}zin}\ and\ \citenamefont
  {De~Dominicis}(1975)}]{brezin1975field}%
  \BibitemOpen
  \bibfield  {author} {\bibinfo {author} {\bibfnamefont {E.}~\bibnamefont
  {Br{\'e}zin}}\ and\ \bibinfo {author} {\bibfnamefont {C.}~\bibnamefont
  {De~Dominicis}},\ }\href@noop {} {\bibfield  {journal} {\bibinfo  {journal}
  {Physical Review B}\ }\textbf {\bibinfo {volume} {12}},\ \bibinfo {pages}
  {4954} (\bibinfo {year} {1975})}\BibitemShut {NoStop}%
\bibitem [{\citenamefont {T{\"a}uber}(1992)}]{tauberdissertation}%
  \BibitemOpen
  \bibfield  {author} {\bibinfo {author} {\bibfnamefont {U.~C.}\ \bibnamefont
  {T{\"a}uber}},\ }\emph {\bibinfo {title} {{Koexistenzanomalien in der Dynamik
  isotroper Systeme}}},\ \href@noop {} {Ph.D. thesis},\ \bibinfo  {school}
  {Technische Universit{\"a}t M{\"u}nchen} (\bibinfo {year} {1992})\BibitemShut
  {NoStop}%
\bibitem [{\citenamefont {De~Dominicis}(1976)}]{dominicis1976technics}%
  \BibitemOpen
  \bibfield  {author} {\bibinfo {author} {\bibfnamefont {C.}~\bibnamefont
  {De~Dominicis}},\ }in\ \href@noop {} {\emph {\bibinfo {booktitle} {J.
  Phys.(Paris), Colloq}}}\ (\bibinfo {year} {1976})\ p.~\bibinfo {pages}
  {C1}\BibitemShut {NoStop}%
\bibitem [{\citenamefont {Lawrie}(1981)}]{lawrie1981goldstone}%
  \BibitemOpen
  \bibfield  {author} {\bibinfo {author} {\bibfnamefont {I.~D.}\ \bibnamefont
  {Lawrie}},\ }\href@noop {} {\bibfield  {journal} {\bibinfo  {journal}
  {Journal of Physics A: Mathematical and General}\ }\textbf {\bibinfo {volume}
  {14}},\ \bibinfo {pages} {2489} (\bibinfo {year} {1981})}\BibitemShut
  {NoStop}%
\bibitem [{\citenamefont {Anderson}(1958)}]{anderson1958random}%
  \BibitemOpen
  \bibfield  {author} {\bibinfo {author} {\bibfnamefont {P.~W.}\ \bibnamefont
  {Anderson}},\ }\href@noop {} {\bibfield  {journal} {\bibinfo  {journal}
  {Physical Review}\ }\textbf {\bibinfo {volume} {112}},\ \bibinfo {pages}
  {1900} (\bibinfo {year} {1958})}\BibitemShut {NoStop}%
\bibitem [{\citenamefont {Br{\'e}zin}\ and\ \citenamefont
  {Wallace}(1973)}]{brezin1973critical}%
  \BibitemOpen
  \bibfield  {author} {\bibinfo {author} {\bibfnamefont {E.}~\bibnamefont
  {Br{\'e}zin}}\ and\ \bibinfo {author} {\bibfnamefont {D.~J.}\ \bibnamefont
  {Wallace}},\ }\href@noop {} {\bibfield  {journal} {\bibinfo  {journal}
  {Physical Review B}\ }\textbf {\bibinfo {volume} {7}},\ \bibinfo {pages}
  {1967} (\bibinfo {year} {1973})}\BibitemShut {NoStop}%
\bibitem [{\citenamefont {Nelson}(1976)}]{nelson1976coexistence}%
  \BibitemOpen
  \bibfield  {author} {\bibinfo {author} {\bibfnamefont {D.~R.}\ \bibnamefont
  {Nelson}},\ }\href@noop {} {\bibfield  {journal} {\bibinfo  {journal}
  {Physical Review B}\ }\textbf {\bibinfo {volume} {13}},\ \bibinfo {pages}
  {2222} (\bibinfo {year} {1976})}\BibitemShut {NoStop}%
\bibitem [{\citenamefont {T{\"a}uber}\ and\ \citenamefont
  {Schwabl}(1992)}]{tauber1992critical}%
  \BibitemOpen
  \bibfield  {author} {\bibinfo {author} {\bibfnamefont {U.~C.}\ \bibnamefont
  {T{\"a}uber}}\ and\ \bibinfo {author} {\bibfnamefont {F.}~\bibnamefont
  {Schwabl}},\ }\href@noop {} {\bibfield  {journal} {\bibinfo  {journal}
  {Physical Review B}\ }\textbf {\bibinfo {volume} {46}},\ \bibinfo {pages}
  {3337} (\bibinfo {year} {1992})}\BibitemShut {NoStop}%
\bibitem [{\citenamefont {Cardy}(1996)}]{cardy1996scaling}%
  \BibitemOpen
  \bibfield  {author} {\bibinfo {author} {\bibfnamefont {J.~L.}\ \bibnamefont
  {Cardy}},\ }\href@noop {} {\emph {\bibinfo {title} {Scaling and
  renormalization in statistical physics}}},\ Vol.~\bibinfo {volume} {5}\
  (\bibinfo  {publisher} {Cambridge University Press},\ \bibinfo {year}
  {1996})\BibitemShut {NoStop}%
\bibitem [{\citenamefont {Zinn-Justin}(2021)}]{zinn2021quantum}%
  \BibitemOpen
  \bibfield  {author} {\bibinfo {author} {\bibfnamefont {J.}~\bibnamefont
  {Zinn-Justin}},\ }\href@noop {} {\emph {\bibinfo {title} {Quantum field
  theory and critical phenomena}}},\ Vol.\ \bibinfo {volume} {171}\ (\bibinfo
  {publisher} {Oxford University Press},\ \bibinfo {year} {2021})\BibitemShut
  {NoStop}%
\bibitem [{\citenamefont {Halperin}\ \emph {et~al.}(1976)\citenamefont
  {Halperin}, \citenamefont {Hohenberg},\ and\ \citenamefont
  {Siggia}}]{halperin1976renormalization}%
  \BibitemOpen
  \bibfield  {author} {\bibinfo {author} {\bibfnamefont {B.~I.}\ \bibnamefont
  {Halperin}}, \bibinfo {author} {\bibfnamefont {P.~C.}\ \bibnamefont
  {Hohenberg}},\ and\ \bibinfo {author} {\bibfnamefont {E.~D.}\ \bibnamefont
  {Siggia}},\ }\href@noop {} {\bibfield  {journal} {\bibinfo  {journal}
  {Physical Review B}\ }\textbf {\bibinfo {volume} {13}},\ \bibinfo {pages}
  {1299} (\bibinfo {year} {1976})}\BibitemShut {NoStop}%
\bibitem [{\citenamefont {Schwabl}\ and\ \citenamefont
  {Michel}(1970)}]{schwabl1970hydrodynamics}%
  \BibitemOpen
  \bibfield  {author} {\bibinfo {author} {\bibfnamefont {F.}~\bibnamefont
  {Schwabl}}\ and\ \bibinfo {author} {\bibfnamefont {K.~H.}\ \bibnamefont
  {Michel}},\ }\href@noop {} {\bibfield  {journal} {\bibinfo  {journal}
  {Physical Review B}\ }\textbf {\bibinfo {volume} {2}},\ \bibinfo {pages}
  {189} (\bibinfo {year} {1970})}\BibitemShut {NoStop}%
\bibitem [{\citenamefont {Akkineni}\ and\ \citenamefont
  {T{\"a}uber}(2004)}]{akkineni2004nonequilibrium}%
  \BibitemOpen
  \bibfield  {author} {\bibinfo {author} {\bibfnamefont {V.~K.}\ \bibnamefont
  {Akkineni}}\ and\ \bibinfo {author} {\bibfnamefont {U.~C.}\ \bibnamefont
  {T{\"a}uber}},\ }\href@noop {} {\bibfield  {journal} {\bibinfo  {journal}
  {Physical Review E}\ }\textbf {\bibinfo {volume} {69}},\ \bibinfo {pages}
  {036113} (\bibinfo {year} {2004})}\BibitemShut {NoStop}%
\bibitem [{\citenamefont {Bausch}\ \emph {et~al.}(1976)\citenamefont {Bausch},
  \citenamefont {Janssen},\ and\ \citenamefont
  {Wagner}}]{bausch1976renormalized}%
  \BibitemOpen
  \bibfield  {author} {\bibinfo {author} {\bibfnamefont {R.}~\bibnamefont
  {Bausch}}, \bibinfo {author} {\bibfnamefont {H.~K.}\ \bibnamefont
  {Janssen}},\ and\ \bibinfo {author} {\bibfnamefont {H.}~\bibnamefont
  {Wagner}},\ }\href@noop {} {\bibfield  {journal} {\bibinfo  {journal}
  {Zeitschrift f{\"u}r Physik B Condensed Matter}\ }\textbf {\bibinfo {volume}
  {24}},\ \bibinfo {pages} {113} (\bibinfo {year} {1976})}\BibitemShut
  {NoStop}%
\bibitem [{\citenamefont {Moshe}\ and\ \citenamefont
  {Zinn-Justin}(2003)}]{moshe2003quantum}%
  \BibitemOpen
  \bibfield  {author} {\bibinfo {author} {\bibfnamefont {M.}~\bibnamefont
  {Moshe}}\ and\ \bibinfo {author} {\bibfnamefont {J.}~\bibnamefont
  {Zinn-Justin}},\ }\href@noop {} {\bibfield  {journal} {\bibinfo  {journal}
  {Physics Reports}\ }\textbf {\bibinfo {volume} {385}},\ \bibinfo {pages} {69}
  (\bibinfo {year} {2003})}\BibitemShut {NoStop}%
\bibitem [{\citenamefont {Eyal}\ \emph {et~al.}(1996)\citenamefont {Eyal},
  \citenamefont {Moshe}, \citenamefont {Nishigaki},\ and\ \citenamefont
  {Zinn-Justin}}]{eyal1996n}%
  \BibitemOpen
  \bibfield  {author} {\bibinfo {author} {\bibfnamefont {G.}~\bibnamefont
  {Eyal}}, \bibinfo {author} {\bibfnamefont {M.}~\bibnamefont {Moshe}},
  \bibinfo {author} {\bibfnamefont {S.}~\bibnamefont {Nishigaki}},\ and\
  \bibinfo {author} {\bibfnamefont {J.}~\bibnamefont {Zinn-Justin}},\
  }\href@noop {} {\bibfield  {journal} {\bibinfo  {journal} {Nuclear Physics
  B}\ }\textbf {\bibinfo {volume} {470}},\ \bibinfo {pages} {369} (\bibinfo
  {year} {1996})}\BibitemShut {NoStop}%
\bibitem [{\citenamefont {T{\"a}uber}\ and\ \citenamefont
  {R{\'a}cz}(1997)}]{tauber1997critical}%
  \BibitemOpen
  \bibfield  {author} {\bibinfo {author} {\bibfnamefont {U.~C.}\ \bibnamefont
  {T{\"a}uber}}\ and\ \bibinfo {author} {\bibfnamefont {Z.}~\bibnamefont
  {R{\'a}cz}},\ }\href@noop {} {\bibfield  {journal} {\bibinfo  {journal}
  {Physical Review E}\ }\textbf {\bibinfo {volume} {55}},\ \bibinfo {pages}
  {4120} (\bibinfo {year} {1997})}\BibitemShut {NoStop}%
\bibitem [{\citenamefont {Nandi}\ and\ \citenamefont
  {T{\"a}uber}(2019)}]{nandi2019nonuniversal}%
  \BibitemOpen
  \bibfield  {author} {\bibinfo {author} {\bibfnamefont {R.}~\bibnamefont
  {Nandi}}\ and\ \bibinfo {author} {\bibfnamefont {U.~C.}\ \bibnamefont
  {T{\"a}uber}},\ }\href@noop {} {\bibfield  {journal} {\bibinfo  {journal}
  {Physical Review B}\ }\textbf {\bibinfo {volume} {99}},\ \bibinfo {pages}
  {064417} (\bibinfo {year} {2019})}\BibitemShut {NoStop}%
\bibitem [{\citenamefont {Cardy}\ and\ \citenamefont
  {Hamber}(1980)}]{cardy1980n}%
  \BibitemOpen
  \bibfield  {author} {\bibinfo {author} {\bibfnamefont {J.~L.}\ \bibnamefont
  {Cardy}}\ and\ \bibinfo {author} {\bibfnamefont {H.~W.}\ \bibnamefont
  {Hamber}},\ }\href@noop {} {\bibfield  {journal} {\bibinfo  {journal}
  {Physical Review Letters}\ }\textbf {\bibinfo {volume} {45}},\ \bibinfo
  {pages} {499} (\bibinfo {year} {1980})}\BibitemShut {NoStop}%
\end{thebibliography}%

\end{document}